\documentclass[aps,twocolumn,letterpaper,notitlepage]{revtex4-1}
\usepackage[T1]{fontenc}
\usepackage[latin9]{inputenc}
\usepackage{amsmath}
\usepackage{amssymb}
\usepackage{stmaryrd}
\usepackage{physics}
\usepackage[mathscr]{euscript}
\usepackage[shortlabels]{enumitem}

\usepackage[usenames,dvipsnames]{xcolor}
\usepackage[hyperindex,pdftex, breaklinks,colorlinks = true,linkcolor = blue,urlcolor=blue,citecolor=blue]{hyperref}

\begin{document}
	
	\title{From Magnetoelectric Response to Optical Activity}
	
	\author{Perry T. Mahon}
	\email{pmahon@physics.utoronto.ca}
	\affiliation{Department of Physics, University of Toronto, Toronto, Ontario M5S 1A7, Canada}
	
	\author{J. E. Sipe}
	\email{sipe@physics.utoronto.ca}
	\affiliation{Department of Physics, University of Toronto, Toronto, Ontario M5S 1A7, Canada}
	
	\date{\today}

\begin{abstract}
We apply a microscopic theory of polarization and magnetization to crystalline insulators at zero temperature and consider the orbital electronic contribution of the linear response to spatially varying, time-dependent electromagnetic fields. The charge and current density expectation values generally depend on both the microscopic polarization and magnetization fields, and on the microscopic free charge and current densities. But contributions from the latter vanish in linear response for the class of insulators we consider. Thus we need only consider the former, which can be decomposed into ``site'' polarization and magnetization fields, from which ``site multipole moments'' can be constructed. Macroscopic polarization and magnetization fields follow, and we identify the relevant contributions to them; for electromagnetic fields varying little over a lattice constant these are the electric and magnetic dipole moments per unit volume, and the electric quadrupole moment per unit volume. A description of optical activity and related magneto-optical phenomena follows from the response of these macroscopic quantities to the electromagnetic field and, while in this paper we work within the independent particle and frozen-ion approximations, both optical rotary dispersion and circular dichroism can be described with this strategy. Earlier expressions describing the magnetoelectric effect are recovered as the zero frequency limit of our more general equations. Since our site quantities are introduced with the use of Wannier functions, the site multipole moments and their macroscopic analogs are generally gauge dependent. However, the resulting macroscopic charge and current densities, together with the optical effects to which they lead, are gauge invariant, as would be physically expected.
\end{abstract}

\maketitle

\section{Introduction}
\label{Section1}

In a material that is \textit{optically active} the plane of polarization of light rotates as the light propagates through the medium; the rotation is associated with a difference in the phase velocities of right- and left-handed circularly polarized light. The frequency dependence of the rotation is called \textit{optical rotary dispersion}, and the associated difference in absorption of light of the different circular polarizations is called \textit{circular dichroism}.

The study of optical activity has a long history. Pasteur was the first to associate it with structural dissymmetry \footnote{Cited in Ref.~\cite{Pasteur}}, and as early as 1928 its first quantum mechanical description was given by Rosenfeld \cite{Rosenfeld}. This phenomenon is most often observed in liquid solutions. The usual solvent, water, is not itself optically active, but the solution is optically active if the symmetry group characterizing the structure of the solute molecules contains no improper rotations. Early theoretical treatments involved models of solute molecules based on at least two coupled oscillators at different sites in each molecule \cite{Kirkwood}, and it was natural to associate optical activity with the variation of the electromagnetic field across the molecule. However, an alternate approach \footnote{See, e.g., van Kranendonk and Sipe \cite{Sipe1976}} is to consider the electric and magnetic multipole moments of each molecule as a whole, and to describe their response to the electromagnetic field and its derivatives at a nominal center of the molecule. Optical activity is then typically associated with the response of the electric dipole moment to both the magnetic field and the symmetrized derivative of the electric field, of the magnetic dipole moment to the electric field, and of the electric quadrupole moment to the electric field. For studies of solutions the last contribution is not relevant in practice, since it vanishes when averaged over all orientations of the solute molecules \cite{Dunn}.

Optical activity can also occur in crystalline materials \cite{Nye} with $\alpha$-quartz perhaps the most familiar example. It can be described with the aid of an effective conductivity tensor \cite{Souza2010}, $\sigma^{il}(\boldsymbol{q},\omega)$, that depends on both the frequency $\omega$ and the wave vector $\boldsymbol{q}$ of the electromagnetic field. This tensor relates the linear response of the macroscopic current density $\boldsymbol{J}^{(1)}(\boldsymbol{q},\omega)$ to the macroscopic electric field $\boldsymbol{E}(\boldsymbol{q},\omega$) that induces it,
\begin{align}
J^{i(1)}(\boldsymbol{q},\omega)=\sigma^{il}(\boldsymbol{q},\omega)E^{l}(\boldsymbol{q},\omega),
\label{eq:bulk}
\end{align}
with superscript indices denoting Cartesian components, which are summed over when repeated. An expansion for small $\boldsymbol{q}$, $\sigma^{il}(\boldsymbol{q},\omega)=\sigma^{il}(\omega)+\sigma^{ilj}(\omega)q^{j}+\ldots$, where $\sigma^{il}(\omega)\equiv\sigma^{il}(\boldsymbol{0},\omega)$ and $\sigma^{ilj}(\omega)\equiv(\partial\sigma^{il}(\boldsymbol{q},\omega)/\partial q^{j})_{\boldsymbol{q}=\boldsymbol{0}}$, gives 
\begin{align}
J^{i(1)}(\boldsymbol{q},\omega)=\sigma^{il}(\omega)E^{l}(\boldsymbol{q},\omega)+\sigma^{ilj}(\omega)E^{l}(\boldsymbol{q},\omega)q^{j}+\ldots
\label{eq:Jexpand}
\end{align}
The first term on the right-hand side, when Fourier transformed to position space, gives the usual long-wavelength response, $J^{i(E)}(\boldsymbol{x},\omega)=\sigma^{il}(\omega)E^{l}(\boldsymbol{x},\omega)$. Using Faraday's law, the second term on the right can be rewritten in terms of the magnetic field and the symmetrized spatial derivative of the electric field, and if $\sigma^{ilj}(\omega)$ is nonvanishing then the medium is optically active. From this perspective, optical activity can arise as one of the consequences of ``spatial dispersion'' \cite{Ginzburg}, when a response tensor such as $\sigma^{il}(\boldsymbol{q},\omega)$ depends on $\boldsymbol{q}$ as well as $\omega$. If time-reversal symmetry holds before the medium is subjected to the electromagnetic field, then the optical activity has been called \textit{natural} \cite{Souza2010}. If time-reversal symmetry is broken, then there are generally additional contributions to $\sigma^{ilj}(\omega)$, and as well a rotation of the plane of polarization of light can result from an asymmetric component of $\sigma^{il}(\omega)$, as $\sigma^{il}(\omega)\neq\sigma^{li}(\omega)$ in general. This latter phenomenon can be thought of as an ``internal'' Faraday effect.

Yet such a general treatment of linear optical properties of media based on $\sigma^{il}(\boldsymbol{q},\omega)$ has its drawbacks. First, when using the minimal coupling Hamiltonian and directly calculating the expectation value of the electronic current density operator, ``artificial divergences'' can arise when the number of bands involved in the calculation is necessarily truncated; sum rules must be employed before such a truncation is performed to avoid these \cite{Cardona,Ghahramani}. Second, although one can attribute different constituents of $\sigma^{ilj}(\omega)$ to the purported response of different multipole moments \cite{Souza2010}, those multipole moments, and the physical insight they carry, do not directly arise in the calculation. And third, the bulk relation (\ref{eq:bulk}) and its expansion (\ref{eq:Jexpand}) give little direction on how to even approximately treat the subtleties that would arise if one considered a finite system and had to be concerned with effects at interfaces.

A strategy that is more physical is certainly available for crystalline systems in the ``molecular crystal limit.'' In this limit we imagine molecules, here with no improper rotations in their symmetry group, positioned at lattice sites with a lattice constant sufficiently large that electrons can be considered essentially ``bound'' to one molecule or another, but still much less than the wavelength of light. Adopting the approach of molecular physics \cite{PZW}, multipole moments can be associated with each molecule and from these one can introduce macroscopic fields $\mathscr{P}_{\text{mol}}^{i}(\boldsymbol{x},t)$, $\mathscr{M}_{\text{mol}}^{i}(\boldsymbol{x},t)$, and $\mathscr{Q}_{\text{mol}}^{ij}(\boldsymbol{x},t)$, describing respectively the electric dipole, the magnetic dipole, and the electric quadrupole moment per unit volume of the ``molecular crystal.'' The macroscopic charge and current densities are then given by 
\begin{align}
\varrho_{\text{mol}}(\boldsymbol{x},t)&=-\boldsymbol{\nabla}\boldsymbol{\cdot}\boldsymbol{P}_{\text{mol}}(\boldsymbol{x},t),\nonumber \\
\boldsymbol{J}_{\text{mol}}(\boldsymbol{x},t)&=\frac{\partial\boldsymbol{P}_{\text{mol}}(\boldsymbol{x},t)}{\partial t}+c\boldsymbol{\nabla}\cross\boldsymbol{M}_{\text{mol}}(\boldsymbol{x},t),\label{eq:Jandrho}
\end{align}
where the polarization and magnetization fields are given by
\begin{align}
 P_{\text{mol}}^{i}(\boldsymbol{x},t)&=\mathscr{P}_{\text{mol}}^{i}(\boldsymbol{x},t)-\frac{\partial\mathscr{Q}_{\text{mol}}^{ij}(\boldsymbol{x},t)}{\partial x^{j}}+\ldots, \nonumber\\
M_{\text{mol}}^{i}(\boldsymbol{x},t)&=\mathscr{M}_{\text{mol}}^{i}(\boldsymbol{x},t)+\ldots,
\label{eq:PandM} 
\end{align}
with ``$\ldots$'' indicating contributions from higher-order multipole moments. Neglecting local field corrections, from the response tensors associated with the multipole moments of the molecules themselves one can then identify bulk linear response tensors $\mathring{\chi}_{E}^{il}(\omega)$, $\mathring{\gamma}^{ijl}(\omega)$, $\mathring{\beta}_{\mathscr{P}}^{il}(\omega)$, $\mathring{\beta}_{\mathscr{M}}^{il}(\omega)$, and $\mathring{\chi}_{\mathscr{Q}}^{ijl}(\omega)$ that relate the multipole moments to the macroscopic electric and magnetic fields,
\begin{align}
\mathscr{P}_{\text{mol}}^{i}(\boldsymbol{x},t)&=\mathscr{P}_{\text{mol}}^{i(0)}+\sum_{\omega}e^{-i\omega t}\big(\mathring{\chi}_{E}^{il}(\omega)E^{l}(\boldsymbol{x},\omega)\nonumber\\
&\qquad+\mathring{\gamma}^{ijl}(\omega)F^{jl}(\boldsymbol{x},\omega)+\mathring{\beta}_{\mathscr{P}}^{il}(\omega)B^{l}(\boldsymbol{x},\omega)+\ldots\big),\nonumber\\
\mathscr{Q}_{\text{mol}}^{ij}(\boldsymbol{x},t)&=\mathscr{Q}_{\text{mol}}^{ij(0)}+\sum_{\omega}e^{-i\omega t}\big(\mathring{\chi}_{\mathscr{Q}}^{ijl}(\omega)E^{l}(\boldsymbol{x},\omega)+\ldots\big),\nonumber \\
\mathscr{M}_{\text{mol}}^{i}(\boldsymbol{x},t)&=\mathscr{M}_{\text{mol}}^{i(0)}+\sum_{\omega}e^{-i\omega t}\big(\mathring{\beta}_{\mathscr{M}}^{il}(\omega)E^{l}(\boldsymbol{x},\omega)+\ldots\big),
\label{eq:linear_response} 
\end{align}
where the superscript $(0)$ identifies the contribution to a net quantity from the unperturbed system, ``$\ldots$'' here indicate contributions that are higher order in the macroscopic electric and magnetic fields and their derivatives, including the linear
response of $\boldsymbol{\mathscr{M}}_{\text{mol}}$ to $\boldsymbol{B}$,
\begin{align*}
F^{jl}(\boldsymbol{x},\omega)\equiv\frac{1}{2}\left(\frac{\partial E^{j}(\boldsymbol{x},\omega)}{\partial x^{l}}+\frac{\partial E^{l}(\boldsymbol{x},\omega)}{\partial x^{j}}\right)
\end{align*}
is the symmetrized (spatial) derivative of the macroscopic electric field evaluated at $\boldsymbol{x}$, and the circle accents identify that these linear response tensors are valid in the molecular crystal limit.

Using (\ref{eq:linear_response},\ref{eq:PandM}) in (\ref{eq:Jandrho}), transforming to wave-vector space, and comparing with (\ref{eq:Jexpand}), we can construct $\sigma_{\text{mol}}^{ilj}(\omega)$ in terms of $\mathring{\gamma}^{ijl}(\omega)$, $\mathring{\beta}_{\mathscr{P}}^{il}(\omega)$, $\mathring{\beta}_{\mathscr{M}}^{il}(\omega)$, and $\mathring{\chi}_{\mathscr{Q}}^{ijl}(\omega)$. Such a calculation based on molecular response, done in terms of the multipole Hamiltonian familiar in molecular physics \cite{Healybook}, does not suffer from the artificial divergences mentioned above; thus the resulting expression for $\sigma_{\text{mol}}^{ilj}(\omega)$ is well behaved. In addition, if time-reversal symmetry is broken before the molecules are subjected to the electromagnetic field, then $\mathring{\chi}_{E}^{il}(\omega)\neq\mathring{\chi}_{E}^{li}(\omega)$, which gives $\sigma^{il}_{\text{mol}}(\omega)\neq\sigma^{li}_{\text{mol}}(\omega)$, leading to another source of the rotation of the plane of polarization of light as it propagates through the molecular crystal. Here the multipole moments of the molecules explicitly appear, and with the underlying macroscopic fields $\mathscr{P}_{\text{mol}}^{i}(\boldsymbol{x},t)$, $\mathscr{M}_{\text{mol}}^{i}(\boldsymbol{x},t)$, and $\mathscr{Q}_{\text{mol}}^{ij}(\boldsymbol{x},t)$ in hand one could begin to consider electrodynamics in the presence of interfaces.

But now what of more realistic models of crystalline materials, wherein the molecular crystal limit is not satisfied? Although there are no centers with which particular electrons are associated, in the ``modern theory of polarization and magnetization'' one can still define electric and magnetic dipole moments \cite{Resta1994,Resta2005,Resta2006}, albeit indirectly, through the response of the electronic charge and current densities to external electromagnetic fields. This approach is generally focused on the limit of static applied fields to insulators, the inclusion of higher-order moments in this framework is work in progress \cite{Hughes2017,Hughes2019}, and its generalization to optical fields is not obvious.

We recently introduced \cite{Mahon2019,Mahon2020} a general approach to calculating both the static and the optical perturbative response of a medium based on the introduction of \textit{microscopic} polarization and magnetization fields, $\boldsymbol{p}(\boldsymbol{x},t)$ and $\boldsymbol{m}(\boldsymbol{x},t)$. The usual macroscopic fields $\boldsymbol{P}(\boldsymbol{x},t)$ and $\boldsymbol{M}(\boldsymbol{x},t)$ are defined as the spatial averages of the corresponding microscopic fields. In general there are also microscopic free charge and current densities, the spatial averages of which are identified as the macroscopic free charge and current densities. However, at zero temperature, for the class of insulating crystals to which we restrict ourselves in this paper -- which includes ordinary insulators \footnote{By ``ordinary insulator'' we mean crystalline insulators supporting Bloch energy eigenstates for which there exists no topological obstruction to choosing a smooth gauge that can respect some underlying symmetry of the system. For instance, there exists no obstruction to choosing a time-reversal or inversion symmetric gauge for a system with the same discrete symmetry.} and $\mathbb{Z}_2$ topological insulators -- those free charge and current densities vanish in linear response, and the full microscopic response, to first order in the electromagnetic field, is captured by $\boldsymbol{p}(\boldsymbol{x},t)$ and $\boldsymbol{m}(\boldsymbol{x},t)$.

With the introduction of Wannier functions, the microscopic fields $\boldsymbol{p}(\boldsymbol{x},t)$ and $\boldsymbol{m}(\boldsymbol{x},t)$ can be decomposed into constituents associated with each lattice site, and these ``site'' contributions can be expanded in terms of a series of ``site multipole moments.'' The spatial average of these microscopic fields then leads to an expansion of the macroscopic polarization and magnetization fields in the form (\ref{eq:PandM}), even if the molecular crystal limit does not hold. Further, the response of the multipole moments associated with each lattice site can be calculated in terms of the electromagnetic field and its derivatives evaluated at that site, and leads naturally to a description of the linear response that follows the form (\ref{eq:linear_response}), again even though the molecular crystal limit does not hold. As well, the artificial divergences that can plague standard minimal coupling calculations are absent.

In this approach the site contributions to the electronic component of the microscopic polarization and magnetization fields, and thus to their multipole moments, depend on a modified form of the Wannier functions resulting from a generalized Peierls substitution \cite{Mahon2019}. There is also a well-known ``gauge freedom'' in choosing the original Wannier functions from which the modified functions are constructed, for they can be altered by adjusting the $\boldsymbol{k}$-dependent unitary transformation relating them to the Bloch energy eigenstates. In general this leads to a ``gauge dependence'' of the site multipole moments, both initially and in their response to the electromagnetic field. And while exponentially localized Wannier functions (ELWFs) would of course be a natural choice for the original Wannier functions, we show that whatever choice is made the resulting electronic charge and current densities predicted are gauge invariant \footnote{The ``gauge freedom'' referred to here only considers transformations amongst the set of initially occupied or unoccupied electronic energy eigenstates.}, as would be physically expected; the expressions we extract for $\sigma^{il}(\omega)$ and $\sigma^{ilj}(\omega)$ are thus gauge invariant. Even within the independent particle and frozen-ion approximations, which we adopt in this work, we believe this is the first derivation of $\sigma^{ilj}(\omega)$ for an insulator that is valid not only at frequencies below the band gap, but also at frequencies above the band gap where absorption can occur. Thus, the $\sigma^{ilj}(\omega)$ we present provides a description for both the optical rotary dispersion and the circular dichroism of crystalline insulators. At frequencies below the band gap we find agreement with an earlier calculation of $\sigma^{ilj}(\omega)$ \cite{Souza2010} that focused on that limit.

The special case of static and uniform electric and magnetic fields is particularly interesting. In that limit the tensor describing the modification of the polarization due to the electric field becomes symmetric, even in the absence of time-reversal symmetry in the unperturbed crystal. But in the absence of both time-reversal and spatial inversion symmetry, a magnetic field can still induce a polarization and an electric field can still induce a magnetization. This phenomenon is called the \textit{magnetoelectric effect} \cite{Fiebig2005}; in an earlier work \cite{Mahon2020} we used our approach to derive the so-called orbital magnetoelectric polarizability (OMP) tensor that describes the magnetoelectric effect in the limit of fixed ion cores and with the neglect of spin contributions, and found agreement with earlier studies based on the ``modern theory of polarization and magnetization'' \cite{Essin2010,Malashevich2010}. Optical activity can be understood as arising from the generalization of the magnetoelectric effect to finite frequencies, where the electromagnetic field is necessarily not uniform; time-reversal symmetry then need not be broken for the phenomenon to occur. And as our calculation is based on a microscopic identification of polarization and magnetization fields, we can identify a finite frequency generalization of the Chern-Simons contribution to the OMP tensor; this contribution is isotropic and thus does not lead to an induced electronic charge-current density in the bulk, which makes it inaccessible to approaches based on the bulk charge-current density response alone.

Finally, since our calculation is based on the identification of site quantities, we can easily compare the general response of a crystal to that of a crystal in the molecular crystal limit mentioned above. In this paper we identify expressions for the response tensors $\chi_{E}^{il}(\omega)$, $\gamma^{ijl}(\omega)$, $\beta_{\mathscr{P}}^{il}(\omega)$, $\beta_{\mathscr{M}}^{il}(\omega)$, and $\chi_{\mathscr{Q}}^{ijl}(\omega)$ in both cases, indicating the response tensors that are valid in the molecular crystal limit by a circle accent as we have above. In particular, while the OMP tensor is identified with $\beta_{\mathscr{P}}^{il}(0)=\beta_{\mathscr{M}}^{li}(0)$, the relation $\mathring{\beta}_{\mathscr{P}}^{il}(\omega)=\mathring{\beta}_{\mathscr{M}}^{li}(-\omega)$ continues to hold for finite frequencies in the molecular crystal limit, but it fails for a crystal more generally. Thus our approach is well positioned to explore the boundary between molecular physics and condensed matter physics in their descriptions of optical activity, and indeed of other optical phenomena.

The structure of this paper is as follows. In Section \ref{Section2} we present the basic expressions for the microscopic polarization and magnetization fields, identify the site multipole moments, and present their relation to the macroscopic response functions; some of the details are relegated to Appendices \ref{Appendix:Relator} and \ref{Appendix:MacroFields}. The linear response of a crystalline insulator, within the independent particle approximation, is presented in Section \ref{Section3}. Here for simplicity we neglect the spin degree of freedom and treat the ion cores as fixed. The response of the site multipole moments is detailed in Section \ref{Section4}, where we also consider some of the symmetries of the response tensors. In Section \ref{Section5} we construct the linear response of the macroscopic charge and current densities, and identify $\sigma^{il}(\omega)$ and $\sigma^{ilj}(\omega)$; their constituent tensors are listed in Appendix \ref{Appendix:Tensors}, and in Appendices \ref{Appendix:InducedCurrent} and \ref{Appendix:InducedCharge} we confirm that the response is gauge invariant and thus that $\sigma^{il}(\omega)$ and $\sigma^{ilj}(\omega)$ are as well. We also consider the special case of frequencies below the band gap of the insulator and confirm, using a result presented in Appendix \ref{Appendix:Conductivity}, that we have agreement with earlier work for $\sigma^{ilj}(\omega)$ \cite{Souza2010}. In Section \ref{Section6} we consider the molecular crystal limit and show that in this limit our general crystalline expressions reduce to what would be expected. We discuss and conclude in Section \ref{Section7}.

\section{Multipole moments}
\label{Section2}

In earlier work \cite{Mahon2019,Mahon2020} we showed how the (total) microscopic charge and current densities can be written as 
\begin{align}
 \rho(\boldsymbol{x},t)&=-\boldsymbol{\nabla}\boldsymbol{\cdot}\boldsymbol{p}(\boldsymbol{x},t)+\rho_{F}(\boldsymbol{x},t),\nonumber \\
 \boldsymbol{j}(\boldsymbol{x},t)&=\frac{\partial\boldsymbol{p}(\boldsymbol{x},t)}{\partial t}+c\boldsymbol{\nabla}\cross\boldsymbol{m}(\boldsymbol{x},t)+\boldsymbol{j}_{F}(\boldsymbol{x},t),\label{eq:micro}
\end{align}
where, in this work,
\begin{align*}
\rho(\boldsymbol{x},t)&\equiv\expval{\hat{\rho}(\boldsymbol{x},t)}+\rho^{\text{ion}}(\boldsymbol{x}),\\
\boldsymbol{j}(\boldsymbol{x},t)&\equiv\expval{\hat{\boldsymbol{j}}(\boldsymbol{x},t)},
\end{align*}
with $\rho^{\text{ion}}(\boldsymbol{x})$ the charge density associated with fixed ion cores, and $\expval{\hat{\rho}(\boldsymbol{x},t)}$ and $\expval{\hat{\boldsymbol{j}}(\boldsymbol{x},t)}$
the expectation values of the microscopic electronic charge and current
density operators, respectively \footnote{In contrast to past work \cite{Mahon2020} we explicitly include contributions from \textit{both} the electronic charge and current densities, and the ionic charge and current (assumed to vanish) densities in the microscopic polarization $\boldsymbol{p}(\boldsymbol{x},t)$ and magnetization $\boldsymbol{m}(\boldsymbol{x},t)$ fields. The sole role played by the static ionic charges, apart from providing the underlying periodic potential of the system, is that, together with the electronic charge density, result in a vanishing $\rho^{(0)}_{F}(\boldsymbol{x},t)$.}. These operators are obtained from the minimal coupling Hamiltonian via Noether's theorem and involve the electron field operators and their adjoint, which we take to be the dynamical degrees of freedom of the crystalline system; they evolve under the minimal coupling Hamiltonian, which results in the (assumed classical) electromagnetic field entering (\ref{eq:micro}), and thus in both $\boldsymbol{p}(\boldsymbol{x},t)$ and $\boldsymbol{m}(\boldsymbol{x},t)$ generally having a nontrivial dependence on time \cite{Mahon2019,Mahon2020}. These microscopic fields can generally be decomposed as a sum of constituent fields \cite{Mahon2019}, one associated with each Bravais lattice vector $\boldsymbol{R}$ characterizing the structure of the unperturbed crystalline system,
\begin{align}
\boldsymbol{p}(\boldsymbol{x},t)&=\sum_{\boldsymbol{R}}\boldsymbol{p}_{\boldsymbol{R}}(\boldsymbol{x},t),\nonumber\\
\boldsymbol{m}(\boldsymbol{x},t)&=\sum_{\boldsymbol{R}}\boldsymbol{m}_{\boldsymbol{R}}(\boldsymbol{x},t).\label{eq:total_pandm} 
\end{align}
Each ``site'' polarization $\boldsymbol{p}_{\boldsymbol{R}}(\boldsymbol{x},t)$
is related to a portion $\rho_{\boldsymbol{R}}(\boldsymbol{x},t)$
of the (total) charge density that is associated with the lattice site $\boldsymbol{R}$, and each ``site'' magnetization $\boldsymbol{m}_{\boldsymbol{R}}(\boldsymbol{x},t)$ is related to a portion $\boldsymbol{j}_{\boldsymbol{R}}(\boldsymbol{x},t)+\tilde{\boldsymbol{j}}_{\boldsymbol{R}}(\boldsymbol{x},t)$ of the electronic current density that is associated with the lattice site $\boldsymbol{R}$,
\begin{align}
p_{\boldsymbol{R}}^{i}(\boldsymbol{x},t)&\equiv\int s^{i}(\boldsymbol{x};\boldsymbol{y},\boldsymbol{R})\rho_{\boldsymbol{R}}(\boldsymbol{y},t)d\boldsymbol{y},\nonumber\\
m_{\boldsymbol{R}}^{i}(\boldsymbol{x},t)&\equiv\frac{1}{c}\int\alpha^{ib}(\boldsymbol{x};\boldsymbol{y},\boldsymbol{R})\Big(j_{\boldsymbol{R}}^{b}(\boldsymbol{y},t)+\tilde{j}_{\boldsymbol{R}}^{b}(\boldsymbol{y},t)\Big)d\boldsymbol{y},\label{eq:site_contributions} 
\end{align}
where the ``relators'' $s^{i}(\boldsymbol{x};\boldsymbol{y},\boldsymbol{R})$ and $\alpha^{ib}(\boldsymbol{x};\boldsymbol{y},\boldsymbol{R})$ have been introduced and discussed previously \cite{Mahon2019}; they are presented in Appendix \ref{Appendix:Relator}. In general the microscopic ``free'' charge and current densities, $\rho_{F}(\boldsymbol{x},t)$ and $\boldsymbol{j}_{F}(\boldsymbol{x},t)$, are also relevant. However, in this paper we assume the crystal to be in its zero temperature ground state before the electromagnetic field is applied and so, for the class of insulators considered here and specified below, both the unperturbed free charge and current densities, and their linear response to the electric and magnetic fields vanish \cite{Mahon2019}. This is as would be expected physically, and we can henceforth neglect those fields.

The macroscopic polarization and magnetization fields, $\boldsymbol{P}(\boldsymbol{x},t)$ and $\boldsymbol{M}(\boldsymbol{x},t)$, can be identified as spatial averages of the microscopic fields (\ref{eq:total_pandm}), as discussed in Appendix \ref{Appendix:MacroFields}. Anticipating the integration over each site contribution (\ref{eq:site_contributions}) associated with such spatial averaging, we perform a formal expansion of each site contribution in terms of Dirac $\delta$ functions and their derivatives about that site, as we detail in Appendix \ref{Appendix:Relator}. The expansions are characterized by their dependence on a parameter $u$, and explicitly retaining the terms that are at most linear in that parameter we find 
\begin{align}
p_{\boldsymbol{R}}^{i}(\boldsymbol{x},t)&=\mu_{\boldsymbol{R}}^{i}(t)\delta(\boldsymbol{x}-\boldsymbol{R})-q_{\boldsymbol{R}}^{ij}(t)\frac{\partial\delta(\boldsymbol{x}-\boldsymbol{R})}{\partial x^{j}}+\ldots,\nonumber\\
m_{\boldsymbol{R}}^{i}(\boldsymbol{x},t)&=\nu_{\boldsymbol{R}}^{i}(t)\delta(\boldsymbol{x}-\boldsymbol{R})+\ldots,
\label{eq:pmexpand} 
\end{align}
where 
\begin{align}
\mu_{\boldsymbol{R}}^{i}(t)\equiv\int\big(y^{i}-R^{i}\big)\rho_{\boldsymbol{R}}(\boldsymbol{y},t)d\boldsymbol{y}\label{eq:ED}
\end{align}
is the electric dipole moment,
\begin{align}
q_{\boldsymbol{R}}^{ij}(t)\equiv\frac{1}{2}\int\big(y^{i}-R^{i}\big)\big(y^{j}-R^{j}\big)\rho_{\boldsymbol{R}}(\boldsymbol{y},t)d\boldsymbol{y}\label{eq:EQ}
\end{align}
is the electric quadrupole moment, and 
\begin{align}
\nu_{\boldsymbol{R}}^{i}(t)\equiv\frac{\epsilon^{iab}}{2c}\int\big(y^{a}-R^{a}\big)\Big(j_{\boldsymbol{R}}^{b}(\boldsymbol{y},t)+\tilde{j}_{\boldsymbol{R}}^{b}(\boldsymbol{y},t)\Big)d\boldsymbol{y}\label{eq:MD}
\end{align}
is the magnetic dipole moment, each associated with lattice site $\boldsymbol{R}$; here $\epsilon^{iab}$ is the Levi-Civita symbol. Terms that are higher order in $u$, indicated by ``$\ldots$'' in the expansions (\ref{eq:pmexpand}), involve the electric octupole moment, the magnetic quadrupole moment, and higher-order moments.

For the sort of systems considered here, within the independent particle approximation one can physically expect the response of the moments $\mu_{\boldsymbol{R}}^{i}(t)$, $q_{\boldsymbol{R}}^{ij}(t)$, and $\nu_{\boldsymbol{R}}^{i}(t)$ to the microscopic electric and magnetic fields to depend on those fields in the neighborhood of $\boldsymbol{R}$. The approximation of neglecting ``local field corrections'', which we adopt here, involves taking those fields to simply be the macroscopic fields $\boldsymbol{E}(\boldsymbol{x},t)$ and $\boldsymbol{B}(\boldsymbol{x},t)$ that are the spatial averages of the microscopic electric and magnetic fields; we call these macroscopic fields the ``Maxwell fields'' (see Appendix \ref{Appendix:MacroFields}). With this approximation, we show in Section \ref{Section4} that the linear response of each site moment (\ref{eq:ED},\ref{eq:EQ},\ref{eq:MD}) can be related to the Maxwell fields evaluated at that site, $\boldsymbol{E}(\boldsymbol{R},t)$ and $\boldsymbol{B}(\boldsymbol{R},t)$, and their spatial derivatives there. Then, implementing the usual Fourier series analysis,
\begin{align}
 g(t)\equiv\sum_{\omega}e^{-i\omega t}g(\omega),
\label{eq:time_frequency}
\end{align}
we find that the relevant terms are
\begin{align}
\mu_{\boldsymbol{R}}^{i}(t)&=\mu_{\boldsymbol{R}}^{i(0)}+\sum_{\omega}e^{-i\omega t}\big(\Omega_{uc}\chi_{E}^{il}(\omega)E^{l}(\boldsymbol{R},\omega) \nonumber\\
&\qquad+\Omega_{uc}\gamma^{ijl}(\omega)F^{jl}(\boldsymbol{R},\omega)\nonumber\\
&\qquad+\Omega_{uc}\beta_{\mathscr{P}}^{il}(\omega)B^{l}(\boldsymbol{R},\omega)+\ldots\big), \nonumber\\
q_{\boldsymbol{R}}^{ij}(t)&=q_{\boldsymbol{R}}^{ij(0)}+\sum_{\omega}e^{-i\omega t}\big(\Omega_{uc}\chi_{\mathscr{Q}}^{ijl}(\omega)E^{l}(\boldsymbol{R},\omega)+\ldots\big), \nonumber \\
\nu_{\boldsymbol{R}}^{i}(t)&=\nu_{\boldsymbol{R}}^{i(0)}+\sum_{\omega}e^{-i\omega t}\big(\Omega_{uc}\beta_{\mathscr{M}}^{il}(\omega)E^{l}(\boldsymbol{R},\omega)+\ldots\big). \label{eq:linear_response-1} 
\end{align}
We have chosen to introduce a unit cell volume $\Omega_{uc}$ here because, with the neglect of local field corrections, the response tensors $\chi_{E}^{il}(\omega)$, $\gamma^{ijl}(\omega)$, $\beta_{\mathscr{P}}^{il}$($\omega)$, $\beta_{\mathscr{M}}^{il}(\omega)$, and $\chi_{\mathscr{Q}}^{ijl}(\omega)$ appearing here reduce to those of (\ref{eq:linear_response}), in the molecular crystal limit. We show in Appendix \ref{Appendix:MacroFields} that macroscopic multipole moments, analogous to those appearing in (\ref{eq:linear_response}), can be constructed from the corresponding site multipole moments (\ref{eq:linear_response-1}) (see (\ref{eq:mac_moments}) and (\ref{eq:proj_result})), such that
\begin{align}
\mathscr{P}^{i}(\boldsymbol{x},t)&=\frac{1}{\Omega_{uc}}\mu^{i(0)}_{\boldsymbol{R}}+\sum_{\omega}e^{-i\omega t}\big(\chi_{E}^{il}(\omega)E^{l}(\boldsymbol{x},\omega)\nonumber\\
&\qquad+\gamma^{ijl}(\omega)F^{jl}(\boldsymbol{x},\omega)+\beta_{\mathscr{P}}^{il}(\omega)B^{l}(\boldsymbol{x},\omega)+\ldots\big), \nonumber\\
\mathscr{Q}^{ij}(\boldsymbol{x},t)&=\frac{1}{\Omega_{uc}}q^{ij(0)}_{\boldsymbol{R}}+\sum_{\omega}e^{-i\omega t}\big(\chi_{\mathscr{Q}}^{ijl}(\omega)E^{l}(\boldsymbol{x},\omega)+\ldots\big), \nonumber \\
\mathscr{M}^{i}(\boldsymbol{x},t)&=\frac{1}{\Omega_{uc}}\nu^{i(0)}_{\boldsymbol{R}}+\sum_{\omega}e^{-i\omega t}\big(\beta_{\mathscr{M}}^{il}(\omega)E^{l}(\boldsymbol{x},\omega)+\ldots\big),
\label{eq:linear_response-2} 
\end{align}
where the unperturbed contributions simply acquire a factor as they are in fact independent of $\boldsymbol{R}$. Further, the macroscopic charge and current densities are given by
\begin{align}
\varrho(\boldsymbol{x},t)&=-\boldsymbol{\nabla}\boldsymbol{\cdot}\boldsymbol{P}(\boldsymbol{x},t),\nonumber \\
\boldsymbol{J}(\boldsymbol{x},t)&=\frac{\partial\boldsymbol{P}(\boldsymbol{x},t)}{\partial t}+c\boldsymbol{\nabla}\cross\boldsymbol{M}(\boldsymbol{x},t),
\label{eq:Jandrho-1}
\end{align}
with macroscopic polarization and magnetization fields
\begin{align}
P^{i}(\boldsymbol{x},t)&=\mathscr{P}^{i}(\boldsymbol{x},t)-\frac{\partial\mathscr{Q}^{ij}(\boldsymbol{x},t)}{\partial x^{j}}+\ldots,\nonumber\\
M^{i}(\boldsymbol{x},t)&=\mathscr{M}^{i}(\boldsymbol{x},t)+\ldots,\label{eq:PandM-1} 
\end{align}
even far from the molecular crystal limit. Notably in the systems we consider here, the unperturbed contributions to (\ref{eq:linear_response-2}) vanish when implemented in (\ref{eq:Jandrho-1}). Hence, the lowest-order charge and current densities arise due to the linear response tensors $\chi_{E}^{il}(\omega)$, $\gamma^{ijl}(\omega)$, $\beta_{\mathscr{P}}^{il}$($\omega)$,
$\beta_{\mathscr{M}}^{il}(\omega)$, and $\chi_{\mathscr{Q}}^{ijl}(\omega)$. In the next two sections we turn to the calculation of these response tensors.

\section{Linear response}
\label{Section3}

The charge and current densities associated with each lattice site that were mentioned above can be written as
\begin{align}
\rho_{\boldsymbol{R}}(\boldsymbol{x},t)&=\sum_{\alpha\beta\boldsymbol{R}'\boldsymbol{R}''}\rho_{\beta\boldsymbol{R}';\alpha\boldsymbol{R}''}(\boldsymbol{x},\boldsymbol{R};t)\eta_{\alpha\boldsymbol{R}'';\beta\boldsymbol{R}'}(t)+\rho_{\boldsymbol{R}}^{\text{ion}}(\boldsymbol{x}),\nonumber\\
\boldsymbol{j}_{\boldsymbol{R}}(\boldsymbol{x},t)&=\sum_{\alpha\beta\boldsymbol{R}'\boldsymbol{R}''}\boldsymbol{j}_{\beta\boldsymbol{R}';\alpha\boldsymbol{R}''}(\boldsymbol{x},\boldsymbol{R};t)\eta_{\alpha\boldsymbol{R}'';\beta\boldsymbol{R}'}(t),\nonumber \\
\tilde{\boldsymbol{j}}_{\boldsymbol{R}}(\boldsymbol{x},t)&=\sum_{\alpha\beta\boldsymbol{R}'\boldsymbol{R}''}\tilde{\boldsymbol{j}}_{\beta\boldsymbol{R}';\alpha\boldsymbol{R}''}(\boldsymbol{x},\boldsymbol{R};t)\eta_{\alpha\boldsymbol{R}'';\beta\boldsymbol{R}'}(t),\label{eq:site_expand} 
\end{align}
where $\rho_{\boldsymbol{R}}^{\text{ion}}(\boldsymbol{x})$ is the static contribution to the charge density associated with lattice site $\boldsymbol{R}$ due to the appropriate ion core(s) and where the $\rho_{\beta\boldsymbol{R}';\alpha\boldsymbol{R}''}(\boldsymbol{x},\boldsymbol{R};t)$,
$\boldsymbol{j}_{\beta\boldsymbol{R}';\alpha\boldsymbol{R}''}(\boldsymbol{x},\boldsymbol{R};t)$, and $\tilde{\boldsymbol{j}}_{\beta\boldsymbol{R}';\alpha\boldsymbol{R}''}(\boldsymbol{x},\boldsymbol{R};t)$ are generalized (electronic) ``site quantity matrix elements'' that have been presented earlier \cite{Mahon2019}. These quantities can be reasonably expected to vanish unless $\boldsymbol{x}$ is ``close'' to $\boldsymbol{R}$, guaranteeing that $\rho_{\boldsymbol{R}}(\boldsymbol{x},t)$, $\boldsymbol{j}_{\boldsymbol{R}}(\boldsymbol{x},t)$, and $\tilde{\boldsymbol{j}}_{\boldsymbol{R}}(\boldsymbol{x},t)$ have that property as well. To avoid possible confusion we note that the total microscopic charge and current densities are given by $\rho(\boldsymbol{x},t)=\sum\limits_{\boldsymbol{R}}\rho_{\boldsymbol{R}}(\boldsymbol{x},t)$ and $\boldsymbol{j}(\boldsymbol{x},t)=\sum\limits_{\boldsymbol{R}}\boldsymbol{j}_{\boldsymbol{R}}(\boldsymbol{x},t)$, while $\sum\limits_{\boldsymbol{R}}\tilde{\boldsymbol{j}}_{\boldsymbol{R}}(\boldsymbol{x},t)\neq\boldsymbol{0}$ in general; that is, $\boldsymbol{p}_{\boldsymbol{R}}(\boldsymbol{x},t)$ depends on the net charge density that is associated with $\boldsymbol{R}$, while $\boldsymbol{m}_{\boldsymbol{R}}(\boldsymbol{x},t)$ is sensitive to only a portion of the current density that is associated with $\boldsymbol{R}$.

It is clear from (\ref{eq:site_expand}) that the single-particle density matrix, $\eta_{\alpha\boldsymbol{R}'';\beta\boldsymbol{R}'}(t)$, is central in the identification of electronic ``site'' quantities and in describing their dynamics \cite{Mahon2019}. This object captures the electronic transition amplitude from a particular Wannier orbital of type $\alpha$ associated with lattice site $\boldsymbol{R}''$ to a Wannier orbital of type $\beta$ associated with $\boldsymbol{R}'$, at time $t$ (see Eq.~(33,36) of Ref.~\cite{Mahon2019}).

\subsection{Dynamical and compositional contributions to the multipole moments}

The site quantities of primary interest are the Cartesian components of the lowest-order multipole moments (\ref{eq:ED},\ref{eq:EQ},\ref{eq:MD}) that are associated with lattice site $\boldsymbol{R}$. Indicating such a site quantity generally by $\Lambda_{\boldsymbol{R}}(t)$, it is clear that upon inserting the relevant term(s) (\ref{eq:site_expand}) in the desired site multipole moment expression (\ref{eq:ED},\ref{eq:EQ},\ref{eq:MD}), $\Lambda_{\boldsymbol{R}}(t)$ is generally of the form
\begin{align}
\Lambda_{\boldsymbol{R}}(t)=\sum_{\alpha\beta\boldsymbol{R}'\boldsymbol{R}''}\Lambda_{\beta\boldsymbol{R}';\alpha\boldsymbol{R}''}(\boldsymbol{R};t)\eta_{\alpha\boldsymbol{R}'';\beta\boldsymbol{R}'}(t)+\Lambda_{\boldsymbol{R}}^{\text{ion}},
\label{lambdaSite}
\end{align}
where $\Lambda_{\beta\boldsymbol{R}';\alpha\boldsymbol{R}''}(\boldsymbol{R};t)$ is a general (electronic) site quantity matrix element and $\Lambda_{\boldsymbol{R}}^{\text{ion}}$ involves $\rho_{\boldsymbol{R}}^{\text{ion}}(\boldsymbol{x})$. In addition to the dependence of the single-particle density matrix on time, which would be expected in the presence of a time-dependent electromagnetic field, the site quantity matrix elements appearing in (\ref{eq:site_expand}) also have a time dependence -- and thus so do the $\Lambda_{\beta\boldsymbol{R}';\alpha\boldsymbol{R}''}(\boldsymbol{R};t)$ associated with the various site multipole moments -- because they themselves depend on the electromagnetic field. This sort of dependence is not unexpected in the response of systems to the full electromagnetic field. The diamagnetic response of an atom, for example, is not due to a change in its wave function when a magnetic field is applied, which would be captured by the single-particle density matrix, but rather arises because the expression of the charge velocity in terms of the canonical momentum is modified.

We begin by expanding all objects in powers of the electromagnetic field, such that
\begin{align*}
\eta_{\alpha\boldsymbol{R}'';\beta\boldsymbol{R}'}(t)&=\eta_{\alpha\boldsymbol{R}'';\beta\boldsymbol{R}'}^{(0)}+\eta_{\alpha\boldsymbol{R}'';\beta\boldsymbol{R}'}^{(1)}(t)+\ldots,\\
\Lambda_{\beta\boldsymbol{R}';\alpha\boldsymbol{R}''}(\boldsymbol{R};t)&=\Lambda_{\beta\boldsymbol{R}';\alpha\boldsymbol{R}''}^{(0)}(\boldsymbol{R})+\Lambda_{\beta\boldsymbol{R}';\alpha\boldsymbol{R}''}^{(1)}(\boldsymbol{R};t)+\ldots,
\end{align*}
etc. Again, the superscript $(0)$ denotes the contribution to the
quantity that is independent of the Maxwell fields; this is the value the object would take in the unperturbed system. The superscript $(1)$ denotes the linear response of the quantity to the Maxwell fields \footnote{In past work we used the terminology ``first-order modification'' rather than ``linear response'' because, in that work, uniform dc electromagnetic fields were considered.}. Here ``$\ldots$'' represent terms that are higher than first order in the Maxwell fields and will later be neglected. Also, for $n\neq 0$, $\rho_{\boldsymbol{R}}^{\text{ion}(n)}(\boldsymbol{x})=0$ and consequently $\Lambda_{\boldsymbol{R}}^{\text{ion}(n)}=0$ as the ion cores are assumed fixed; thus, in describing the electronic response, the net response of the system is captured. From (\ref{lambdaSite}) it is clear that there are two (electronic) contributions to the linear response of a general site quantity to the Maxwell fields,
\begin{align}
\Lambda_{\boldsymbol{R}}^{(1)}(t)=\Lambda_{\boldsymbol{R}}^{(1;\text{I})}(t)+\Lambda_{\boldsymbol{R}}^{(1;\text{II})}(t).
\label{lambdaLinear}
\end{align}
We have called \cite{Mahon2020} the first term on the right-hand side,
\begin{align}
\Lambda_{\boldsymbol{R}}^{(1;\text{I})}(t)\equiv\sum_{\alpha\beta\boldsymbol{R}'\boldsymbol{R}''}\Lambda_{\beta\boldsymbol{R}';\alpha\boldsymbol{R}''}^{(0)}(\boldsymbol{R})\eta_{\alpha\boldsymbol{R}'';\beta\boldsymbol{R}'}^{(1)}(t),\label{eq:dynamical_lambda}
\end{align}
a ``dynamical'' contribution to the linear response, because it arises from modifications to the unperturbed single-particle density matrix due to the Maxwell fields, and the other term,
\begin{align}
\Lambda_{\boldsymbol{R}}^{(1;\text{II})}(t)\equiv\sum_{\alpha\beta\boldsymbol{R}'\boldsymbol{R}''}\Lambda_{\beta\boldsymbol{R}';\alpha\boldsymbol{R}''}^{(1)}(\boldsymbol{R};t)\eta_{\alpha\boldsymbol{R}'';\beta\boldsymbol{R}'}^{(0)},\label{eq:compositional_lambda}
\end{align}
a ``compositional'' contribution, because it arises due to the way in which the site quantity matrix elements themselves depend on the Maxwell fields. As we will show, (\ref{eq:compositional_lambda}) only describes first-order modifications of single-site properties as a result of the electromagnetic field. 

Moreover, we will generally decompose the linear response of a site quantity (\ref{lambdaLinear}) as a sum of the contributions from the Maxwell electric field, its symmeterized derivative, the Maxwell magnetic field, and higher-order derivatives of these fields, such that
\begin{align}
\Lambda_{\boldsymbol{R}}^{(1)}(t)=\Lambda_{\boldsymbol{R}}^{(E)}(t)+\Lambda_{\boldsymbol{R}}^{(F)}(t)+\Lambda_{\boldsymbol{R}}^{(B)}(t)+\ldots
\label{lambdaDecomp}
\end{align}
In general each of the constituents on the right-hand side of (\ref{lambdaDecomp}) is composed of a dynamical contribution and a compositional contribution; for instance,
\begin{align*}
\Lambda_{\boldsymbol{R}}^{(E)}(t)=\Lambda_{\boldsymbol{R}}^{(E;\text{I})}(t)+\Lambda_{\boldsymbol{R}}^{(E;\text{II})}(t).
\end{align*}
However, for each site multipole moment that is considered only a limited number of the constituents in (\ref{lambdaDecomp}) are retained; this is detailed in Section \ref{Section4}.

In the remainder of this section we determine the evolution of $\eta_{\alpha\boldsymbol{R}'';\beta\boldsymbol{R}'}^{(1)}(t)$ from the initial $\eta_{\alpha\boldsymbol{R}'';\beta\boldsymbol{R}'}^{(0)}$, and in the following section we combine those results with the $\Lambda_{\beta\boldsymbol{R}';\alpha\boldsymbol{R}''}^{(0)}(\boldsymbol{R},t)$ and $\Lambda_{\beta\boldsymbol{R}';\alpha\boldsymbol{R}''}^{(1)}(\boldsymbol{R},t)$ appropriately, to find the linear response of the site multipole moments.

\subsection{Evolution of the single-particle density matrix}

In the independent particle approximation, the equations of motion governing the evolution of the (electronic) single-particle density matrix elements take the form \cite{Mahon2019}
\begin{align}
i\hbar\frac{\partial\eta_{\alpha\boldsymbol{R}'';\beta\boldsymbol{R}'}(t)}{\partial t}=\sum_{\mu\nu\boldsymbol{R}_{1}\boldsymbol{R}_{2}}\mathfrak{F}_{\alpha\boldsymbol{R}'';\beta\boldsymbol{R}'}^{\mu\boldsymbol{R}_{1};\nu\boldsymbol{R}_{2}}(t)\eta_{\mu\boldsymbol{R}_{1};\nu\boldsymbol{R}_{2}}(t),\label{EDM-eom}
\end{align}
where
\begin{align*}
\mathfrak{F}_{\alpha\boldsymbol{R}'';\beta\boldsymbol{R}'}^{\mu\boldsymbol{R}_{1};\nu\boldsymbol{R}_{2}}(t)&=\delta_{\nu\beta}\delta_{\boldsymbol{R}_{2}\boldsymbol{R}'}e^{i\Delta(\boldsymbol{R}'',\boldsymbol{R}_{\text{a}},\boldsymbol{R}_{1},\boldsymbol{R}';t)}\bar{H}_{\alpha\boldsymbol{R}'';\mu\boldsymbol{R}_{1}}(\boldsymbol{R}_{\text{a}},t)\nonumber \\
&-\delta_{\mu\alpha}\delta_{\boldsymbol{R}_{1}\boldsymbol{R}''}e^{i\Delta(\boldsymbol{R}'',\boldsymbol{R}_{2},\boldsymbol{R}_{\text{a}},\boldsymbol{R}';t)}\bar{H}_{\nu\boldsymbol{R}_{2};\beta\boldsymbol{R}'}(\boldsymbol{R}_{\text{a}},t)\nonumber \\
&-\hbar\frac{\partial\Delta(\boldsymbol{R}'',\boldsymbol{R}_{\text{a}},\boldsymbol{R}';t)}{\partial t}\delta_{\nu\beta}\delta_{\mu\alpha}\delta_{\boldsymbol{R}_{2}\boldsymbol{R}'}\delta_{\boldsymbol{R}_{1}\boldsymbol{R}''}.
%\label{eq:Fscript}
\end{align*}
The quantities $\bar{H}_{\mu\boldsymbol{R}_{1};\nu\boldsymbol{R}_{2}}(\boldsymbol{R}_{\text{a}},t)$ can be understood as generalized ``hopping'' matrix elements and are as previously defined \cite{Mahon2020}. With the neglect of local field corrections they involve the Maxwell fields in the neighborhood of the lattice sites appearing, including lattice site $\boldsymbol{R}_{\text{a}}$. This lattice site can be arbitrarily chosen \cite{Mahon2020}, and we discuss its choice below. The Maxwell field $\boldsymbol{B}(\boldsymbol{x},t)$ also enters in the quantities $\Delta(\boldsymbol{R}'',\ldots,\boldsymbol{R}';t)$, which are proportional to the magnetic flux through the surface generated by connecting the points $(\boldsymbol{R}'',\ldots,\boldsymbol{R}')$ with straight lines, when the usual choice of straight-line paths for the relators is adopted (see Appendix \ref{Appendix:Relator}). In this work, this choice is always made.

An expansion of the hopping matrix elements in powers of the electromagnetic field \cite{Mahon2020} gives 
\begin{align*}
\bar{H}_{\mu\boldsymbol{R}_{1};\nu\boldsymbol{R}_{2}}(\boldsymbol{R}_{\text{a}},t)=H_{\mu\boldsymbol{R}_{1};\nu\boldsymbol{R}_{2}}^{(0)}+\bar{H}_{\mu\boldsymbol{R}_{1};\nu\boldsymbol{R}_{2}}^{(1)}(\boldsymbol{R}_{\text{a}},t)+\ldots,
%\label{eq:Hbar_expansion}
\end{align*}
with 
\begin{align}
{H}_{\mu\boldsymbol{R}_{1};\nu\boldsymbol{R}_{2}}^{(0)}=\int W_{\mu\boldsymbol{R}_{1}}^{*}(\boldsymbol{x})H_{0}\big(\boldsymbol{x},\boldsymbol{\mathfrak{p}}(\boldsymbol{x})\big)W_{\nu\boldsymbol{R}_{2}}(\boldsymbol{x})d\boldsymbol{x},\label{eq:Hnought}
\end{align}
where $W_{\alpha\boldsymbol{R}}(\boldsymbol{x})\equiv\braket{\boldsymbol{x}}{\alpha\boldsymbol{R}}$ is the ELWF identified by its type $\alpha$ and the lattice site $\boldsymbol{R}$ with which it is associated, and $H_{0}\big(\boldsymbol{x},\boldsymbol{\mathfrak{p}}(\boldsymbol{x})\big)$ is the differential operator that governs the dynamics of the electron field operators in the unperturbed infinite crystal; we take 
\begin{align*}
\boldsymbol{\mathfrak{p}}(\boldsymbol{x})=\frac{\hbar}{i}\boldsymbol{\nabla}-\frac{e}{c}\boldsymbol{A}_{\text{static}}(\boldsymbol{x}),
\end{align*}
where we allow for a static and periodic magnetic field described by a vector potential satisfying $\boldsymbol{A}_{\text{static}}(\boldsymbol{x})=\boldsymbol{A}_{\text{static}}(\boldsymbol{x}+\boldsymbol{R})$ for any lattice vector $\boldsymbol{R}$ \footnote{The inclusion of such an ``internal'' static magnetic field allows the unperturbed Hamiltonian to generally break time-reversal symmetry, which leads to the possibility of ``internal'' magneto-electric effects.}. The eigenfunctions of $H_{0}\big(\boldsymbol{x},\boldsymbol{\mathfrak{p}}(\boldsymbol{x})\big)$ are of the usual Bloch form $\psi_{n\boldsymbol{k}}(\boldsymbol{x})\equiv\braket{\boldsymbol{x}}{\psi_{n\boldsymbol{k}}}=e^{i\boldsymbol{k}\boldsymbol{\cdot}\boldsymbol{x}}u_{n\boldsymbol{k}}(\boldsymbol{x})/(2\pi)^{3/2}$ with $u_{n\boldsymbol{k}}(\boldsymbol{x})\equiv\braket{\boldsymbol{x}}{n\boldsymbol{k}}$ being a cell-periodic function, and are identified by a band index $n$ and an index $\boldsymbol{k}$ identifying the associated crystal momentum $\hbar\boldsymbol{k}$; we denote the corresponding eigenvalues by $E_{n\boldsymbol{k}}$. These energy eigenfunctions can be used to construct ELWFs \cite{Brouder,Vanderbilt2011,Marzari2012,Panati,Troyer2016} via
\begin{align}
\braket{\boldsymbol{x}}{\alpha\boldsymbol{R}}=\sqrt{\Omega_{uc}}\int_{\text{BZ}}\frac{d\boldsymbol{k}}{(2\pi)^{3}}e^{i\boldsymbol{k}\boldsymbol{\cdot}(\boldsymbol{x}-\boldsymbol{R})}\braket{\boldsymbol{x}}{\alpha\boldsymbol{k}},\label{WF}
\end{align}
where the vectors $\ket{\alpha\boldsymbol{k}}$ are related to the vectors $\ket{n\boldsymbol{k}}$ by a (unitary) ``multiband gauge transformation'',
\begin{align}
\ket{\alpha\boldsymbol{k}} =\sum_{n}U_{n\alpha}(\boldsymbol{k})\ket{n\boldsymbol{k}}.
\label{gaugeTransf}
\end{align}
Generally for an insulating crystal in its zero temperature ground state there is a filling factor $f_n$ associated with each $\ket{n\boldsymbol{k}}$ that is either $0$ or $1$. And in this paper we restrict ourselves to the class of insulators characterized by the property that the sets of occupied and unoccupied cell-periodic functions $\braket{\boldsymbol{x}}{n\boldsymbol{k}}$ can be used separately to construct sets of ELWFs; this class contains both ordinary insulators and $\mathbb{Z}_{2}$ topological insulators \cite{Vanderbilt2011,Troyer2016}. Thus we can associate an analogous filling factor $f_{\alpha}$ with each $\ket{\alpha\boldsymbol{k}}$ that is also either $0$ or $1$ depending on the occupancy of the $\ket{n\boldsymbol{k}}$ used in the construction of that particular $\ket{\alpha\boldsymbol{k}}$, and so $U_{n\alpha}(\boldsymbol{k})\neq0$ only if $f_n=f_{\alpha}$. That is, (\ref{gaugeTransf}) is a unitary transformation between elements of the (un)occupied subspace of the electronic Hilbert space alone. Associated with the set of vectors $\{\ket{n\boldsymbol{k}}\}$ is a non-Abelian Berry connection,
\begin{align*}
\xi_{mn}^{a}(\boldsymbol{k})&=i\left(m\boldsymbol{k}|\partial_{a}n\boldsymbol{k}\right)\equiv\frac{i}{\Omega_{uc}}\int_{\Omega_{uc}}u_{m\boldsymbol{k}}^{*}(\boldsymbol{x})\frac{\partial u_{n\boldsymbol{k}}(\boldsymbol{x})}{\partial k^{a}}d\boldsymbol{x},
\end{align*}
and with the set of vectors $\{\ket{\alpha\boldsymbol{k}}\}$
is another, 
\begin{align*}
\tilde{\xi}_{\beta\alpha}^{a}(\boldsymbol{k})=i\left(\beta\boldsymbol{k}|\partial_{a}\alpha\boldsymbol{k}\right).
\end{align*}
These objects are related via
\begin{align}
\sum_{\alpha\beta}U_{m\beta}(\boldsymbol{k})\tilde{\xi}_{\beta\alpha}^{a}(\boldsymbol{k})U_{\alpha n}^{\dagger}(\boldsymbol{k})=\xi_{mn}^{a}(\boldsymbol{k})+\mathcal{W}_{mn}^{a}(\boldsymbol{k}),\label{connection}
\end{align}
where we have defined the Hermitian matrix $\mathcal{W}^{a}(\boldsymbol{k})$, populated
by elements 
\begin{align}
\mathcal{W}_{mn}^{a}(\boldsymbol{k})\equiv i\sum\limits _{\alpha}\big(\partial_{a}U_{m\alpha}(\boldsymbol{k})\big)U_{\alpha n}^{\dagger}(\boldsymbol{k}),\label{W}
\end{align}
and in general we adopt the shorthand $\partial_{a}\equiv\partial/\partial k^{a}$. For the class of insulators we consider, $\mathcal{W}_{mn}^{a}(\boldsymbol{k})$ is nonzero only if $f_{m}=f_{n}$. In what follows, the $\boldsymbol{k}$ dependence of the above introduced objects is usually kept implicit.

At zeroth order in the Maxwell fields, the unperturbed expression (\ref{eq:Hnought}) can be implemented into (\ref{EDM-eom}); with this, the elements of the unperturbed single-particle density matrix for the zero temperature ground state are identified to be
\begin{align}
\eta_{\alpha\boldsymbol{R}'';\beta\boldsymbol{R}'}^{(0)}=f_{\alpha}\delta_{\alpha\beta}\delta_{\boldsymbol{R}''\boldsymbol{R}'}.\label{EDMunpert}
\end{align}
Following the same procedure, but now retaining only the terms that are first order in the Maxwell fields, the linear response of the single-particle density matrix is identified as \cite{Mahon2020}
\begin{widetext}
\begin{align}
\eta_{\alpha\boldsymbol{R}'';\beta\boldsymbol{R}'}^{(1)}(\omega)&= -\sum_{\mu\nu\boldsymbol{R}_{1}\boldsymbol{R}_{2}}\sum_{mn}f_{nm}\int_{\text{BZ}}d\boldsymbol{k}d\boldsymbol{k}'\frac{\braket{\alpha\boldsymbol{R}''}{\psi_{m\boldsymbol{k}}}\braket{\psi_{m\boldsymbol{k}}}{\mu\boldsymbol{R}_{1}} H_{\mu\boldsymbol{R}_{1};\nu\boldsymbol{R}_{2}}^{(1)}(\boldsymbol{R}_{\text{a}},\omega)\braket{\nu\boldsymbol{R}_{2}}{\psi_{n\boldsymbol{k}'}}\braket{\psi_{n\boldsymbol{k}'}}{\beta\boldsymbol{R}'}}{E_{m\boldsymbol{k}}-E_{n\boldsymbol{k}'}-\hbar(\omega+i0^{+})}\nonumber\\
&+\frac{i}{2}f_{\beta\alpha}\int W_{\alpha\boldsymbol{R}''}^{*}(\boldsymbol{x})\Big(\Delta(\boldsymbol{R}'',\boldsymbol{x},\boldsymbol{R}_{\text{a}};\omega)+\Delta(\boldsymbol{R}',\boldsymbol{x},\boldsymbol{R}_{\text{a}};\omega)\Big)W_{\beta\boldsymbol{R}'}(\boldsymbol{x})d\boldsymbol{x}, \label{EDM1}
\end{align}
\end{widetext}
(recall (\ref{eq:time_frequency})), where $f_{nm}\equiv f_{n}-f_{m}$,
$f_{\beta\alpha}\equiv f_{\beta}-f_{\alpha}$, and
\begin{align}
H_{\mu\boldsymbol{R}_{1};\nu\boldsymbol{R}_{2}}^{(1)}(\boldsymbol{R}_{\text{a}},\omega)
\equiv\int W_{\mu\boldsymbol{R}_{1}}^{*}(\boldsymbol{x})\mathcal{H}_{\boldsymbol{R}_{\text{a}}}^{(1)}(\boldsymbol{x},\omega)W_{\nu\boldsymbol{R}_{2}}(\boldsymbol{x})d\boldsymbol{x}.\label{Hcorr}
\end{align}
Here $\mathcal{H}_{\boldsymbol{R}_{\text{a}}}^{(1)}(\boldsymbol{x},\omega)$ involves the electromagnetic field via the scalar quantity $\Omega_{\boldsymbol{R}_{\text{a}}}^{0}(\boldsymbol{x},\omega)$ and the vector quantity $\boldsymbol{\Omega}_{\boldsymbol{R}_{\text{a}}}(\boldsymbol{x},\omega).$ Very generally, $\Omega_{\boldsymbol{y}}^{0}(\boldsymbol{x},\omega)$ involves a line integral involving $\boldsymbol{E}(\boldsymbol{z},\omega)$ from $\boldsymbol{y}$ to $\boldsymbol{x}$, while $\boldsymbol{\Omega}_{\boldsymbol{y}}(\boldsymbol{x},\omega)$ involves a more complicated line integral involving $\boldsymbol{B}(\boldsymbol{z},\omega)$ from $\boldsymbol{y}$ to $\boldsymbol{x}$ \cite{Mahon2019}, which also appears in (\ref{EDM1}). For $|\boldsymbol{x}-\boldsymbol{y}|$ on the order of a lattice constant, an expansion \cite{Mahon2020} gives
\begin{align}
\Omega^{a}_{\boldsymbol{y}}(\boldsymbol{x},\omega)&=\frac{1}{2}\epsilon^{alb}B^{l}(\boldsymbol{y},\omega)\big(x^b-y^b\big)+\ldots,\label{eq:Omega}\\
\Omega_{\boldsymbol{y}}^{0}(\boldsymbol{x},\omega)&=\big(x^{l}-y^{l}\big)E^{l}(\boldsymbol{y},\omega)\nonumber\\
&+\frac{1}{2}\big(x^{j}-y^{j}\big)\big(x^{l}-y^{l}\big)F^{jl}(\boldsymbol{y},\omega)+\ldots,
\label{eq:omega0}
\end{align}
and similarly for $|\boldsymbol{x}-\boldsymbol{y}|$ and $|\boldsymbol{x}-\boldsymbol{z}|$ on the order of a lattice constant we have 
\begin{align}
\Delta(\boldsymbol{z},\boldsymbol{x},\boldsymbol{y};\omega)=-\frac{e}{2\hbar c}\epsilon^{lab}B^l(\boldsymbol{y},\omega)\big(z^a-y^a\big)\big(x^b-y^b\big)+\ldots\label{eq:Delta_expand}
\end{align}
(see Appendix \ref{Appendix:Relator}). With the approximations (\ref{eq:Omega},\ref{eq:omega0}) we find we can write \cite{Mahon2020}
\begin{align}
\mathcal{H}_{\boldsymbol{R}_{\text{a}}}^{(1)}(\boldsymbol{x},\omega)=&-e\big(x^{l}-R_{\text{a}}^{l}\big)E^{l}(\boldsymbol{R}_{\text{a}},\omega)\nonumber\\
&-\frac{e}{2}\big(x^{j}-R_{\text{a}}^{j}\big)\big(x^{l}-R_{\text{a}}^{l}\big)F^{jl}(\boldsymbol{R}_{\text{a}},\omega)\nonumber\\
&+\frac{e}{2mc}\epsilon^{lab}B^{l}(\boldsymbol{R}_{\text{a}},\omega)\big(x^b-R_{\text{a}}^b\big)\mathfrak{p}^{a}(\boldsymbol{x})+\ldots
\label{Hcal1}
\end{align}
Thus $\boldsymbol{R}_{\text{a}}$ acts as a natural point about which to expand the electromagnetic field, and a natural choice of $\boldsymbol{R}_{\text{a}}$ for use in (\ref{EDM1}) would be a site ``close'' to $\boldsymbol{R}'$ or $\boldsymbol{R}''$. Still leaving that choice open, we implement (\ref{Hcal1}) in (\ref{EDM1}) to identify the contributions to $\eta_{\alpha\boldsymbol{R}'';\beta\boldsymbol{R}'}^{(1)}(\omega)$ from the electric field, the symmetrized derivative of the electric field, and the magnetic field. We write this decomposition as
\begin{align}
\eta_{\alpha\boldsymbol{R}'';\beta\boldsymbol{R}'}^{(1)}(\omega)&=\eta_{\alpha\boldsymbol{R}'';\beta\boldsymbol{R}'}^{(E)}(\omega)+\eta_{\alpha\boldsymbol{R}'';\beta\boldsymbol{R}'}^{(F)}(\omega)\nonumber\\
&+\eta_{\alpha\boldsymbol{R}'';\beta\boldsymbol{R}'}^{(B)}(\omega)+\ldots,
\label{EDMlinear}
\end{align}
which will allow for the identification of the dynamical contributions to the constituents of (\ref{lambdaDecomp}). Notably we will neglect contributions related to the spatial variation of the magnetic field, since we identify any such terms as higher-order modifications (see Appendix \ref{Appendix:Relator}).

\begin{widetext}
\subsection{Linear response of the single-particle density matrix}

By implementing the first and second terms of (\ref{Hcal1}) in (\ref{EDM1}) via (\ref{Hcorr}), and noting (\ref{eq:Delta_expand}) is independent of $E^{l}(\boldsymbol{x},\omega)$ and $F^{jl}(\boldsymbol{x},\omega)$, the linear response of the single-particle density matrix to the Maxwell electric field and its symmetrized derivative are found to be
\begin{align}
\eta_{\alpha\boldsymbol{R}'';\beta\boldsymbol{R}'}^{(E)}(\omega)=e\Omega_{uc}E^l(\boldsymbol{R}_{\text{a}},\omega)\sum_{mn}f_{nm}\int_{\text{BZ}}\frac{d\boldsymbol{k}}{(2\pi)^3}\frac{e^{i\boldsymbol{k}\boldsymbol{\cdot}(\boldsymbol{R}''-\boldsymbol{R}')}U^\dagger_{\alpha m}\xi^l_{mn}U_{n\beta}}{E_{m\boldsymbol{k}}-E_{n\boldsymbol{k}}-\hbar(\omega+i0^+)}, \label{EDMe}
\end{align}
and 
\begin{align}
\eta_{\alpha\boldsymbol{R}'';\beta\boldsymbol{R}'}^{(F)}(\omega)
&=\frac{e\Omega_{uc}}{2}F^{jl}(\boldsymbol{R}_{\text{a}},\omega)\sum_{mn}f_{nm}\int_{\text{BZ}}\frac{d\boldsymbol{k}}{(2\pi)^3}e^{i\boldsymbol{k}\boldsymbol{\cdot}(\boldsymbol{R}''-\boldsymbol{R}')}U^\dagger_{\alpha m}\mathscr{F}^{jl}_{mn}(\boldsymbol{k},\omega)U_{n\beta} \nonumber \\
&+\frac{ie\Omega_{uc}}{2}F^{jl}(\boldsymbol{R}_{\text{a}},\omega)\sum_{mn}f_{nm}\int_{\text{BZ}}\frac{d\boldsymbol{k}}{(2\pi)^3}\frac{e^{i\boldsymbol{k}\boldsymbol{\cdot}(\boldsymbol{R}''-\boldsymbol{R}')}\xi^l_{mn}}{E_{m\boldsymbol{k}}-E_{n\boldsymbol{k}}-\hbar(\omega+i0^+)}\Big\{U^\dagger_{\alpha m}\big(\partial_j U_{n\beta}\big)-\big(\partial_j U^\dagger_{\alpha m}\big)U_{n \beta}\Big\} \nonumber \\
&+\frac{e\Omega_{uc}}{2}F^{jl}(\boldsymbol{R}_{\text{a}},\omega)\sum_{mn}f_{nm}\int_{\text{BZ}}\frac{d\boldsymbol{k}}{(2\pi)^3}\frac{e^{i\boldsymbol{k}\boldsymbol{\cdot}(\boldsymbol{R}''-\boldsymbol{R}')}U^\dagger_{\alpha m}\xi^l_{mn}U_{n\beta}}{E_{m\boldsymbol{k}}-E_{n\boldsymbol{k}}-\hbar(\omega+i0^+)}\Big\{\big(R''^j-R^j_{\text{a}}\big)+\big(R'^j-R^j_{\text{a}}\big)\Big\},
\label{EDMf}
\end{align}
where we have introduced
\begin{align}
{\mathscr{F}}^{jl}_{mn}(\boldsymbol{k},\omega)&\equiv\sum_{s}\frac{\xi^j_{ms}\xi^l_{sn}}{E_{m\boldsymbol{k}}-E_{n\boldsymbol{k}}-\hbar(\omega+i0^+)}+i\frac{\partial_j(E_{m\boldsymbol{k}}+E_{n\boldsymbol{k}})}{\big(E_{m\boldsymbol{k}}-E_{n\boldsymbol{k}}-\hbar(\omega+i0^+)\big)^2}\xi^l_{mn}. \label{scrF}
\end{align}
The linear response to the Maxwell magnetic field involves the third term of (\ref{Hcal1}), and using (\ref{eq:Delta_expand}) it is found to be
\begin{align}
\eta_{\alpha\boldsymbol{R}'';\beta\boldsymbol{R}'}^{(B)}(\omega)&=\frac{e\Omega_{uc}}{4\hbar c}\epsilon^{lab}B^l(\boldsymbol{R}_{\text{a}},\omega)\sum_{mn}f_{nm}\int_{\text{BZ}}\frac{d\boldsymbol{k}}{(2\pi)^{3}}e^{i\boldsymbol{k}\boldsymbol{\cdot}(\boldsymbol{R}''-\boldsymbol{R}')}U^\dagger_{\alpha m}{\mathscr{B}}^{ab}_{mn}(\boldsymbol{k},\omega)U_{n\beta} \nonumber \\
&+\frac{e\Omega_{uc}}{4\hbar c}\epsilon^{lab}B^l(\boldsymbol{R}_{\text{a}},\omega)\sum_{mn}f_{nm}\int_{\text{BZ}}\frac{d\boldsymbol{k}}{(2\pi)^{3}}\frac{e^{i\boldsymbol{k}\boldsymbol{\cdot}(\boldsymbol{R}''-\boldsymbol{R}')}\big(E_{m\boldsymbol{k}}-E_{n\boldsymbol{k}}\big)\xi^b_{mn}}{E_{m\boldsymbol{k}}-E_{n\boldsymbol{k}}-\hbar(\omega+i0^+)}\Big\{\big(\partial_a U^\dagger_{\alpha m}\big)U_{n \beta}-U^\dagger_{\alpha m}\big(\partial_a U_{n\beta}\big)\Big\} \nonumber \\ 
&-\frac{i\omega e\Omega_{uc}}{4 c}\epsilon^{lab}B^l(\boldsymbol{R}_{\text{a}},\omega)\sum_{mn}f_{nm}\int_{\text{BZ}}\frac{d\boldsymbol{k}}{(2\pi)^{3}}\frac{e^{i\boldsymbol{k}\boldsymbol{\cdot}(\boldsymbol{R}''-\boldsymbol{R}')}U^\dagger_{\alpha m}\xi^a_{mn}U_{n\beta}}{E_{m\boldsymbol{k}}-E_{n\boldsymbol{k}}-\hbar(\omega+i0^+)}\Big\{\big(R''^b-R^b_{\text{a}}\big)+\big(R'^b-R^b_{\text{a}}\big)\Big\}, \label{EDMb2}
\end{align}
where we have introduced
\begin{align}
\mathscr{B}^{ab}_{mn}(\boldsymbol{k},\omega)&\equiv i\sum_{s}\left\{\frac{E_{s\boldsymbol{k}}-E_{n\boldsymbol{k}}}{E_{m\boldsymbol{k}}-E_{n\boldsymbol{k}}-\hbar(\omega+i0^+)}\xi^a_{ms}\xi^b_{sn}+\frac{E_{s\boldsymbol{k}}-E_{m\boldsymbol{k}}}{E_{m\boldsymbol{k}}-E_{n\boldsymbol{k}}-\hbar(\omega+i0^+)}\xi^a_{ms}\xi^b_{sn}
\right\} \nonumber \\
&-\left(2+\frac{\hbar\omega}{E_{m\boldsymbol{k}}-E_{n\boldsymbol{k}}-\hbar(\omega+i0^+)}\right)\frac{\partial_a(E_{m\boldsymbol{k}}+E_{n\boldsymbol{k}})}{E_{m\boldsymbol{k}}-E_{n\boldsymbol{k}}-\hbar(\omega+i0^+)}\xi^b_{mn}.
\label{scrB}
\end{align}
\end{widetext}
In the limit of uniform dc Maxwell fields, both (\ref{EDMf}) and the final term of (\ref{EDMb2}) vanish trivially, and $\mathscr{B}_{mn}^{ab}(\boldsymbol{k},\omega=0)$ reduces to the previously defined $\mathscr{B}_{mn}^{ab}(\boldsymbol{k})$; the above expressions are thus consistent with past results \cite{Mahon2020}. Also, (\ref{EDMe},\ref{EDMf},\ref{EDMb2}) are written as single Brillouin zone integrals. In past work \cite{Mahon2020} we showed explicitly how this reduction to a single $\boldsymbol{k}$-integral arises when implementing (\ref{EDM1}) to find the $\omega=0$ component of (\ref{EDMe}). However, this reduction emerges more generally as a consequence of expressing the variation of the electromagnetic field over the unit cell through the expansion, following from (\ref{eq:Omega},\ref{eq:omega0},\ref{eq:Delta_expand}), in powers of the length of the unit cell divided by the wavelength of light. Upon implementing the resulting expressions in (\ref{EDM1}), via (\ref{Hcorr},\ref{Hcal1}), and using previously introduced identities \cite{Mahon2020}, the reduction to a single $\boldsymbol{k}$-integral occurs.

\section{The response tensors}
\label{Section4}

The linear response of the single-particle density matrix (\ref{EDMlinear}) allows for the identification of the dynamical contributions (\ref{eq:dynamical_lambda}) to the linear response of the site multipole moments (\ref{eq:ED},\ref{eq:EQ},\ref{eq:MD}) to the Maxwell fields and their derivatives. For the compositional contributions (\ref{eq:compositional_lambda}), we implement (\ref{EDMunpert}) and (\ref{eq:time_frequency}) to write
\begin{align*}
\Lambda^{(1;\text{II})}_{\boldsymbol{R}}(\omega)=\sum_{\alpha\boldsymbol{R}'}f_{\alpha}\Lambda_{\alpha\boldsymbol{R}';\alpha\boldsymbol{R}'}^{(1)}(\boldsymbol{R};\omega),
\end{align*}
which can also be decomposed into contributions due to the Maxwell fields and their derivatives. The decomposition of the net linear response is given by (\ref{lambdaDecomp}). However, for a given site multipole moment of interest, $\mu^{i}_{\boldsymbol{R}}(\omega)$, $q^{ij}_{\boldsymbol{R}}(\omega)$, or $\nu^{i}_{\boldsymbol{R}}(\omega)$, we consider \textit{only} those constituents of the associated $\Lambda_{\alpha\boldsymbol{R}';\alpha\boldsymbol{R}'}^{(1)}(\boldsymbol{R};\omega)$ and of $\eta_{\alpha\boldsymbol{R}'';\beta\boldsymbol{R}'}^{(1)}(\omega)$ that lead to the explicitly included first-order terms in (\ref{eq:linear_response-1}); these are $\mu^{i(E)}_{\boldsymbol{R}}(\omega)$, $\mu^{i(F)}_{\boldsymbol{R}}(\omega)$, $\mu^{i(B)}_{\boldsymbol{R}}(\omega)$, $q^{ij(E)}_{\boldsymbol{R}}(\omega)$, and $\nu^{i(E)}_{\boldsymbol{R}}(\omega)$. We have justified the retention of only these terms in Appendix \ref{Appendix:Relator}. From (\ref{eq:ED},\ref{eq:EQ},\ref{eq:MD}) follow the relevant site quantity matrix elements associated with the site multipole moments of interest in terms of the site quantity matrix elements appearing in (\ref{eq:site_expand}). These have been presented earlier \cite{Mahon2019}, and we now use them to determine the desired response tensors.

We are finally in a position to set $\boldsymbol{R}_{\text{a}}$. When considering the dynamical contribution to the linear response of a particular multipole moment associated with lattice site $\boldsymbol{R}$ to a particular Maxwell field or its derivative, we always choose $\boldsymbol{R}_{\text{a}}=\boldsymbol{R}$ in the constituent of $\eta_{\alpha\boldsymbol{R}'';\beta\boldsymbol{R}'}^{(1)}(\omega)$ being implemented. For instance, when considering $\boldsymbol{\mu}^{(E;\text{I})}_{\boldsymbol{R}}(\omega)$, we set $\boldsymbol{R}_{\text{a}}=\boldsymbol{R}$ in $\eta_{\alpha\boldsymbol{R}'';\beta\boldsymbol{R}'}^{(E)}(\omega)$. In the expressions that follow, one of the matrix element indices $\boldsymbol{R}'$ or $\boldsymbol{R}''$ always equals $\boldsymbol{R}$, making the use of the expansions (\ref{eq:Omega},\ref{eq:omega0},\ref{eq:Delta_expand}) in deriving (\ref{EDMe},\ref{EDMf},\ref{EDMb2}) sensible. As well, we are always able to manipulate the expressions for the $\Lambda_{\alpha\boldsymbol{R}';\alpha\boldsymbol{R}'}^{(1)}(\boldsymbol{R};\omega)$ that appear in such a way that the Maxwell fields are evaluated at $\boldsymbol{R}$. Collectively, this results in the net linear response of the moments associated with $\boldsymbol{R}$ being related to the electric field, the magnetic field, and the symmetrized derivative of the electric field evaluated at $\boldsymbol{R}$, and facilitates the passage to a relation between the macroscopic polarization and magnetization and the Maxwell fields (see (\ref{eq:linear_response-1},\ref{eq:linear_response-2}) and Appendix \ref{Appendix:MacroFields}).
\begin{widetext}
\subsection{Linear response of the electric moments}

\subsubsection{Dipole response to the electric field}

We begin with the linear response of a site electric dipole moment to the Maxwell electric field, $\boldsymbol{\mu}_{\boldsymbol{R}}^{(E)}(\omega)$. While $\rho_{\alpha\boldsymbol{R}';\alpha\boldsymbol{R}'}^{(1)}(\boldsymbol{x},\boldsymbol{R};t)$ depends on the magnetic field, it does not depend on the electric field; the compositional contribution $\mu^{i(E;\text{II})}_{\boldsymbol{R}}(\omega)$ vanishes. The linear response of this quantity to the electric field is thus entirely dynamical -- it is solely due to $\eta_{\alpha\boldsymbol{R}'';\beta\boldsymbol{R}'}^{(E)}(\omega)$ -- and is given by 
\begin{align}
\mu_{\boldsymbol{R}}^{i{(E)}}(\omega)&=\sum\limits _{\alpha\beta\boldsymbol{R}'\boldsymbol{R}''}\left[\int\big(y^{i}-R^{i}\big)\rho_{\beta\boldsymbol{R}';\alpha\boldsymbol{R}''}^{(0)}(\boldsymbol{y},\boldsymbol{R})d\boldsymbol{y}\right]\eta_{\alpha\boldsymbol{R}'';\beta\boldsymbol{R}'}^{(E)}(\omega)\nonumber \\
& =e^{2}\Omega_{uc}E^{l}(\boldsymbol{R},\omega)\sum_{mn}f_{nm}\int_{\text{BZ}}\frac{d\boldsymbol{k}}{(2\pi)^{3}}\frac{\xi_{mn}^{l}\xi_{nm}^{i}}{E_{m\boldsymbol{k}}-E_{n\boldsymbol{k}}-\hbar(\omega+i0^{+})}.
\label{muE}
\end{align}
From (\ref{eq:linear_response-1}) we identify 
\begin{align}
\chi_{E}^{il}(\omega)=e^{2}\sum_{mn}f_{nm}\int_{\text{BZ}}\frac{d\boldsymbol{k}}{(2\pi)^{3}}\frac{\xi_{mn}^{l}\xi_{nm}^{i}}{E_{m\boldsymbol{k}}-E_{n\boldsymbol{k}}-\hbar(\omega+i0^{+})},
\label{chiE}
\end{align}
which is gauge invariant in that it is independent of $U_{n\alpha}(\boldsymbol{k})$. In general (\ref{chiE}) is not symmetric under exchange of Cartesian components $i$ and $l$. However, if the unperturbed crystal is time-reversal symmetric, then one can show $\chi_{E}^{il}(\omega)$ is equal to $\chi_{E}^{li}(\omega)$; in the $\omega\rightarrow 0$ limit the exchange of these indices is always symmetric, and if absorption is neglected, then $\chi_{E}^{il}(\omega)$ is equal to $\chi_{E}^{li}(-\omega)$ \footnote{The asymmetry of this response tensor is not unexpected; see, e.g., Chen \textit{et al.}~\cite{Chen2019}}. To obtain (\ref{muE}) we have implemented previously introduced \cite{Mahon2020} identities, and in the remainder of this section we often do so; Eq.~(8,14,15) of that work are particularly relevant.

We now take into account the spatial variation of the Maxwell electric field. The compositional contribution vanishes, as in the response calculated above; the linear response is entirely dynamical, and it is given by
\begin{align}
\mu_{\boldsymbol{R}}^{i{(F)}}(\omega)&=\sum\limits _{\alpha\beta\boldsymbol{R}'\boldsymbol{R}''}\left[\int\big(y^{i}-R^{i}\big)\rho_{\beta\boldsymbol{R}';\alpha\boldsymbol{R}''}^{(0)}(\boldsymbol{y},\boldsymbol{R})d\boldsymbol{y}\right]\eta_{\alpha\boldsymbol{R}'';\beta\boldsymbol{R}'}^{(F)}(\omega)\nonumber \\
&=\frac{e^{2}\Omega_{uc}}{2}F^{jl}(\boldsymbol{R},\omega)\sum_{mn}f_{nm}\int_{\text{BZ}}\frac{d\boldsymbol{k}}{(2\pi)^{3}}\Bigg\{\mathscr{F}_{mn}^{jl}(\boldsymbol{k},\omega)\xi_{nm}^{i}+\sum_{s}\frac{\xi_{mn}^{l}\mathcal{W}_{ns}^{j}\xi_{sm}^{i}+\xi_{ns}^{i}\mathcal{W}_{sm}^{j}\xi_{mn}^{l}}{E_{m\boldsymbol{k}}-E_{n\boldsymbol{k}}-\hbar(\omega+i0^{+})}\Bigg\}.\label{muF}
\end{align}
The two distinct contributions appearing in the braces of (\ref{muF}) originate individually from the first and second lines of (\ref{EDMf}), while the contribution from final line of (\ref{EDMf}) vanishes. Via (\ref{eq:linear_response-1}) we again identify the relevant response tensor. We explicitly symmeterize the indices labeling Cartesian components that are contracted with the symmeterized derivative of the Maxwell electric field, and we find
\begin{align}
\gamma^{ijl}(\omega)&=\frac{e^{2}}{4}\sum_{mn}f_{nm}\int_{\text{BZ}}\frac{d\boldsymbol{k}}{(2\pi)^{3}}\Bigg\{\Big(\mathscr{F}_{mn}^{jl}(\boldsymbol{k},\omega)+\mathscr{F}_{mn}^{lj}(\boldsymbol{k},\omega)\Big)\xi_{nm}^{i}\nonumber\\
&\quad\qquad\qquad\qquad\qquad\qquad+\sum\limits _{s}\frac{(\xi_{mn}^{l}\mathcal{W}_{ns}^{j}+\xi_{mn}^{j}\mathcal{W}_{ns}^{l})\xi_{sm}^{i}+\xi_{ns}^{i}(\mathcal{W}_{sm}^{j}\xi_{mn}^{l}+\mathcal{W}_{sm}^{l}\xi_{mn}^{j})}{E_{m\boldsymbol{k}}-E_{n\boldsymbol{k}}-\hbar(\omega+i0^{+})}\Bigg\}.\label{gamma}
\end{align}
Notably $\gamma^{ijl}(\omega)=\gamma^{ilj}(\omega)$; the underlying presence of this symmetry -- even if the indices $j$ and $l$ were not contracted with an object symmetric in those indices, $F^{jl}(\boldsymbol{x},\omega)$, in (\ref{muF}) -- can be recognized by identifying that the objects carrying these indices in (\ref{gamma}) originate from the second term of (\ref{Hcal1}), which was used in (\ref{EDM1}) to obtain (\ref{EDMf}). There $j$ and $l$ are clearly symmetric as the components of $\boldsymbol{x}$ and $\boldsymbol{R}_{\text{a}}$ commute. Unlike $\chi_{E}^{il}(\omega)$, $\gamma^{ijl}(\omega)$ is gauge dependent.

\subsubsection{Dipole response to the magnetic field}

As $\rho_{\alpha\boldsymbol{R}';\alpha\boldsymbol{R}'}^{(1)}(\boldsymbol{x},\boldsymbol{R};t)$
does depend on the magnetic field, there are nonvanishing compositional and dynamical contributions to $\boldsymbol{\mu}_{\boldsymbol{R}}^{(B)}(\omega)$. Letting $\rho_{\beta\boldsymbol{R}';\alpha\boldsymbol{R}''}^{(B)}(\boldsymbol{x},\boldsymbol{R};\omega)$
be the part of $\rho_{\beta\boldsymbol{R}';\alpha\boldsymbol{R}''}^{(1)}(\boldsymbol{x},\boldsymbol{R};\omega)$
that is proportional to the magnetic field, the compositional contribution is given by
\begin{align}
\mu_{\boldsymbol{R}}^{i{(B;\text{II})}}(\omega)&=\sum\limits _{\alpha\boldsymbol{R}'}f_{\alpha}\left[\int\big(y^{i}-R^{i}\big)\rho_{\alpha\boldsymbol{R}';\alpha\boldsymbol{R}'}^{(B)}(\boldsymbol{y},\boldsymbol{R};\omega)d\boldsymbol{y}\right]\nonumber \\
& =\frac{e^{2}\Omega_{uc}}{2\hbar c}\epsilon^{lab}B^{l}(\boldsymbol{R},\omega)\sum_{\alpha\gamma}f_{\alpha}\int_{\text{BZ}}\frac{d\boldsymbol{k}}{(2\pi)^{3}}\text{Re}\Big[\tilde{\xi}_{\alpha\gamma}^{i}\partial_{b}\tilde{\xi}_{\gamma\alpha}^{a}\Big].\label{muBb}
\end{align}
Note that, in going from the first to the final equality, we ensure (using Eq.~(29) of Ref.~\cite{Mahon2019}) the $\Delta(\boldsymbol{R}_1,\boldsymbol{y},\boldsymbol{R};\omega)$ that enters via $\rho_{\alpha\boldsymbol{R}';\alpha\boldsymbol{R}'}^{(B)}(\boldsymbol{y},\boldsymbol{R};\omega)$ (see Eq.~(28,45) of Ref.~\cite{Mahon2019}) is in a form that, upon implementing (\ref{eq:Delta_expand}), the magnetic field is evaluated at $\boldsymbol{R}$. Writing $\beta_{\mathscr{P}}^{il(\text{II})}$ as the compositional contribution to $\beta_{\mathscr{P}}^{il}(\omega)$ (see (\ref{eq:linear_response-2})), we have
\begin{align}
& \beta_{\mathscr{P}}^{il(\text{II})}=\frac{e^{2}}{2\hbar c}\epsilon^{lab}\sum_{\alpha\gamma}f_{\alpha}\int_{\text{BZ}}\frac{d\boldsymbol{k}}{(2\pi)^{3}}\text{Re}\Big[\tilde{\xi}_{\alpha\gamma}^{i}\partial_{b}\tilde{\xi}_{\gamma\alpha}^{a}\Big], \label{eq:betaII}
\end{align}
which is again gauge dependent. Interestingly, $\beta_{\mathscr{P}}^{il(\text{II})}$ is independent of frequency and is identical to the compositional contribution to the tensor describing the linear response of the electric dipole moment to a uniform dc magnetic field (see Ref.~\cite{Mahon2020}).

The form of the dynamical contribution is similar to (\ref{muE}) but with $\eta_{\alpha\boldsymbol{R}'';\beta\boldsymbol{R}'}^{(E)}(\omega)$ replaced by $\eta_{\alpha\boldsymbol{R}'';\beta\boldsymbol{R}'}^{(B)}(\omega)$; denoting its contribution to $\beta_{\mathscr{P}}^{il}(\omega)$ by $\beta_{\mathscr{P}}^{il(\text{I})}(\omega)$, we find 
\begin{align}
& \beta_{\mathscr{P}}^{il(\text{I})}(\omega)=\frac{e^{2}}{4\hbar c}\epsilon^{lab}\sum_{mn}f_{nm}\int_{\text{BZ}}\frac{d\boldsymbol{k}}{(2\pi)^{3}}\Bigg\{\mathscr{B}_{mn}^{ab}(\boldsymbol{k},\omega)\xi_{nm}^{i}+i\frac{\big(E_{m\boldsymbol{k}}-E_{n\boldsymbol{k}}\big)\xi_{mn}^{b}}{E_{m\boldsymbol{k}}-E_{n\boldsymbol{k}}-\hbar(\omega+i0^{+})}\sum_{s}\Big(\xi_{ns}^{i}\mathcal{W}_{sm}^{a}+\mathcal{W}_{ns}^{a}\xi_{sm}^{i}\Big)\Bigg\},\label{muBa}
\end{align}
which is also gauge dependent. The two distinct terms appearing in the braces of (\ref{muBa}) originate individually from the first two lines of (\ref{EDMb2}), and the contribution from final line of (\ref{EDMb2}) vanishes. We separate out the frequency-independent terms that appear in (\ref{muBa}) and combine them with (\ref{eq:betaII}). Together these terms give rise to the previously found \cite{Essin2010,Malashevich2010,Mahon2020} OMP tensor, $\alpha^{il}=\alpha_{\text{G}}^{il}+\delta^{il}\alpha_{\text{CS}}$, where $\alpha_{\text{CS}}$ is termed the Chern-Simons contribution and $\alpha_{\text{G}}^{il}$ the cross-gap contribution; the expressions for these are given in Appendix \ref{Appendix:Tensors}. The remaining terms are used in the construction of an explicitly frequency-dependent response tensor, $\alpha_{\mathscr{P}}^{il}(\omega)$, which vanishes in the $\omega\rightarrow0$ limit. In all, then, we find
\begin{align}
\beta_{\mathscr{P}}^{il}(\omega)&\equiv\beta_{\mathscr{P}}^{il(\text{I})}(\omega)+\beta_{\mathscr{P}}^{il(\text{II})}=\alpha^{il}+\alpha_{\mathscr{P}}^{il}(\omega),\label{eq:betaP}
\end{align}
where we have defined
\begin{align}
\alpha_{\mathscr{P}}^{il}(\omega)&\equiv\frac{e^{2}}{4c}\epsilon^{lab}\sum_{mn}f_{nm}\int_{\text{BZ}}\frac{d\boldsymbol{k}}{(2\pi)^{3}}\frac{\omega}{E_{m\boldsymbol{k}}-E_{n\boldsymbol{k}}-\hbar(\omega+i0^{+})}\Bigg\{\acute{\mathscr{B}}_{mn}^{ab}(\boldsymbol{k},\omega)\xi_{nm}^{i}+i\xi_{mn}^{b}\sum_{s}\Big(\xi_{ns}^{i}\mathcal{W}_{sm}^{a}+\mathcal{W}_{ns}^{a}\xi_{sm}^{i}\Big)\Bigg\},\label{alphaP}\\
\acute{\mathscr{B}}_{mn}^{ab}(\boldsymbol{k},\omega)&\equiv i\sum_{s}\left\{ \frac{E_{s\boldsymbol{k}}-E_{n\boldsymbol{k}}}{E_{m\boldsymbol{k}}-E_{n\boldsymbol{k}}}\xi_{ms}^{a}\xi_{sn}^{b}+\frac{E_{s\boldsymbol{k}}-E_{m\boldsymbol{k}}}{E_{m\boldsymbol{k}}-E_{n\boldsymbol{k}}}\xi_{ms}^{a}\xi_{sn}^{b}\right\} -\left(3+\frac{\hbar\omega}{E_{m\boldsymbol{k}}-E_{n\boldsymbol{k}}-\hbar(\omega+i0^{+})}\right)\frac{\partial_{a}(E_{m\boldsymbol{k}}+E_{n\boldsymbol{k}})}{E_{m\boldsymbol{k}}-E_{n\boldsymbol{k}}}\xi_{mn}^{b}.\label{breveB}
\end{align}
Notably $\alpha_{\mathscr{P}}^{il}(\omega)$ arises due to the linear response of the single-particle density matrix $\eta^{(B)}_{\alpha\boldsymbol{R}'';\beta\boldsymbol{R}'}(\omega)$ \textit{alone}, making it entirely the result of a dynamical contribution. Here $\alpha_{\text{CS}}$ and $\alpha_{\mathscr{P}}^{il}(\omega)$ are gauge dependent, but $\alpha_{\text{G}}^{il}$ is not.

\subsubsection{Quadrupole response to the electric field}

The compositional contribution to $q_{\boldsymbol{R}}^{ij(E)}(\omega)$ vanishes as $\rho_{\alpha\boldsymbol{R}';\alpha\boldsymbol{R}'}^{(1)}(\boldsymbol{x},\boldsymbol{R};t)$ does not depend on the electric field. The dynamical contribution involves $\eta_{\alpha\boldsymbol{R}'';\beta\boldsymbol{R}'}^{(E)}(\omega)$ and again takes the form of (\ref{muE}), except that it will be the second moment (see (\ref{eq:EQ})) of $\rho_{\beta\boldsymbol{R}';\alpha\boldsymbol{R}''}^{(0)}(\boldsymbol{y},\boldsymbol{R})$ that will appear rather than the first. Using the expression for $q_{\boldsymbol{R}}^{ij(E)}(\omega)$ that results, from the second of (\ref{eq:linear_response-1}) we identify
\begin{align}
\chi_{\mathscr{Q}}^{ijl}(\omega)=\frac{e^2}{4}\sum_{mns}f_{nm}\int_{\text{BZ}}\frac{d\boldsymbol{k}}{(2\pi)^3}\frac{\xi^l_{mn}\big(\xi^i_{ns}\xi^j_{sm}+\xi^i_{ns}\mathcal{W}^j_{sm}+\mathcal{W}^i_{ns}\xi^j_{sm}\big)+\xi^l_{mn}\big(\xi^j_{ns}\xi^i_{sm}+\xi^j_{ns}\mathcal{W}^i_{sm}+\mathcal{W}^j_{ns}\xi^i_{sm}\big)}{E_{m\boldsymbol{k}}-E_{n\boldsymbol{k}}-\hbar(\omega+i0^+)},
\label{chiQ}
\end{align}
another gauge-dependent response tensor, with $\chi_{\mathscr{\mathscr{Q}}}^{ijl}(\omega)=\chi_{\mathscr{Q}}^{jil}(\omega)$. This symmetry of the response tensor is a consequence of the symmetry in the definition (\ref{eq:EQ}) of $q_{\boldsymbol{R}}^{ij}(t)$. Notably, both $\chi_{\mathscr{Q}}^{ijl}(\omega)$ and $\gamma^{ijl}(\omega)$ arise from dynamical contributions alone, and are of similar form apart from an energy derivative term that appears in $\gamma^{ijl}(\omega)$.

\subsection{Linear response of the magnetic dipole moment to the electric field}

The expression (\ref{eq:MD}) for a site magnetic dipole moment shows
that there are two contributions; an ``atomic-like'' contribution arising due to $\boldsymbol{j}_{\boldsymbol{R}}(\boldsymbol{y},t)$, and an ``itinerant'' contribution arising due to $\tilde{\boldsymbol{j}}_{\boldsymbol{R}}(\boldsymbol{y},t)$ \cite{Mahon2019}. We denote
the contribution of the first of these to the linear response of the site magnetic dipole
moment to the Maxwell electric field $\boldsymbol{\nu}^{(E)}_{\boldsymbol{R}}(t)$ by $\bar{\boldsymbol{\nu}}^{(E)}_{\boldsymbol{R}}(t)$,
and the second by $\tilde{\boldsymbol{\nu}}^{(E)}_{\boldsymbol{R}}(t)$;
we denote the corresponding contributions to the response tensor $\beta_{\mathscr{M}}^{il}(\omega)$
(recall (\ref{eq:linear_response-1})) by $\bar{\beta}_{\mathscr{M}}^{il}(\omega)$
and $\tilde{\beta}_{\mathscr{M}}^{il}(\omega)$,
\begin{align} \beta_{\mathscr{M}}^{il}(\omega)\equiv\bar{\beta}_{\mathscr{M}}^{il}(\omega)+\tilde{\beta}_{\mathscr{M}}^{il}(\omega).\label{eq:beta_total}
\end{align}
We now identify these contributions.

\subsubsection{Response of the atomic-like contribution}

As $\boldsymbol{j}_{\alpha\boldsymbol{R}';\alpha\boldsymbol{R}'}(\boldsymbol{y},\boldsymbol{R};t)$ does not depend on the electric field, $\bar{\nu}^{i(E;\text{II})}_{\boldsymbol{R}}(t)=0$;
this contribution is entirely dynamical and follows from $\eta_{\alpha\boldsymbol{R}'';\beta\boldsymbol{R}'}^{(E)}(\omega)$. From the resulting expression for $\bar{\nu}^{i(E)}_{\boldsymbol{R}}(\omega)$, we compare to (\ref{eq:linear_response-1}) and extract
\begin{align}
\bar{\beta}_{\mathscr{M}}^{il}(\omega) & =\frac{e^{2}}{4\hbar c}\epsilon^{iab}\sum_{mn}f_{nm}\int_{\text{BZ}}\frac{d\boldsymbol{k}}{(2\pi)^{3}}\left(1+\frac{\hbar\omega}{E_{m\boldsymbol{k}}-E_{n\boldsymbol{k}}-\hbar(\omega+i0^{+})}\right)\nonumber \\
& \quad\qquad\qquad\times\Bigg\{\frac{\partial_{b}(E_{m\boldsymbol{k}}+E_{n\boldsymbol{k}})}{E_{m\boldsymbol{k}}-E_{n\boldsymbol{k}}}\xi_{nm}^{a}\xi_{mn}^{l}+i\sum_{s}\frac{E_{s\boldsymbol{k}}-E_{m\boldsymbol{k}}}{E_{m\boldsymbol{k}}-E_{n\boldsymbol{k}}}\xi_{mn}^{l}\xi_{ns}^{a}\xi_{sm}^{b}+i\sum_{s}\frac{E_{n\boldsymbol{k}}-E_{s\boldsymbol{k}}}{E_{m\boldsymbol{k}}-E_{n\boldsymbol{k}}}\xi_{ns}^{b}\xi_{sm}^{a}\xi_{mn}^{l}\nonumber \\
& \quad\qquad\qquad\qquad+i\sum_{s}\left(\frac{E_{s\boldsymbol{k}}-E_{n\boldsymbol{k}}}{E_{m\boldsymbol{k}}-E_{n\boldsymbol{k}}}-1\right)\xi_{mn}^{l}\mathcal{W}_{ns}^{a}\xi_{sm}^{b}+i\sum_{s}\left(\frac{E_{m\boldsymbol{k}}-E_{s\boldsymbol{k}}}{E_{m\boldsymbol{k}}-E_{n\boldsymbol{k}}}-1\right)\xi_{ns}^{b}\mathcal{W}_{sm}^{a}\xi_{mn}^{l}\Bigg\}.\label{nuBarE}
\end{align}

\subsubsection{Response of the itinerant contribution}

In contrast, since $\tilde{\boldsymbol{j}}_{\alpha\boldsymbol{R}';\alpha\boldsymbol{R}'}(\boldsymbol{y},\boldsymbol{R};t)$ does depend on the electric field, there will be a nonvanishing compositional contribution to $\tilde{\nu}^{i(E)}_{\boldsymbol{R}}(t)$, as well as a dynamical contribution arising from $\eta_{\alpha\boldsymbol{R}'';\beta\boldsymbol{R}'}^{(E)}(\omega)$. We denote the corresponding dynamical contribution to $\tilde{\beta}_{\mathscr{M}}^{il}(\omega)$ by $\tilde{\beta}_{\mathscr{M}}^{il(\text{I})}(\omega)$ and the compositional contribution by $\tilde{\beta}_{\mathscr{M}}^{il(\text{II})}$,
\begin{align*}
\tilde{\beta}_{\mathscr{M}}^{il}(\omega)\equiv\tilde{\beta}_{\mathscr{M}}^{il(\text{I})}(\omega)+\tilde{\beta}_{\mathscr{M}}^{il(\text{II})}.
\end{align*}
We find the compositional contribution to be
\begin{align}
\tilde{\beta}_{\mathscr{M}}^{il(\text{II})}=\frac{e^{2}}{2\hbar c}\epsilon^{iab}\sum_{\alpha\gamma}f_{\alpha}\int_{\text{BZ}}\frac{d\boldsymbol{k}}{(2\pi)^{3}}\text{Re}\big[\tilde{\xi}_{\alpha\gamma}^{l}\partial_{b}\tilde{\xi}_{\gamma\alpha}^{a}\big],\label{nuTildeEb}
\end{align}
which, like (\ref{eq:betaII}), does not depend on frequency. To ensure that the electric field is evaluated at $\boldsymbol{R}$, in reaching (\ref{nuTildeEb}) we have used the form of $\mathfrak{F}_{\alpha\boldsymbol{R}'';\beta\boldsymbol{R}'}^{\mu\boldsymbol{R}_{1};\nu\boldsymbol{R}_{2}}(t)$ presented above in the expression for $\tilde{\boldsymbol{j}}_{\alpha\boldsymbol{R}';\alpha\boldsymbol{R}'}(\boldsymbol{y},\boldsymbol{R};t)$ (Eq.~(60,61,62) of Ref.~\cite{Mahon2019}), and set $\boldsymbol{R}_{\text{a}}=\boldsymbol{R}$. The dynamical contribution is
\begin{align}
\tilde{\beta}_{\mathscr{M}}^{il(\text{I})}(\omega) & =\frac{e^{2}}{4\hbar c}\epsilon^{iab}\sum_{mn}f_{nm}\int_{\text{BZ}}\frac{d\boldsymbol{k}}{(2\pi)^{3}}\left(1+\frac{\hbar\omega}{E_{m\boldsymbol{k}}-E_{n\boldsymbol{k}}-\hbar(\omega+i0^{+})}\right)\nonumber \\
&\quad\qquad\times\Bigg\{\frac{\partial_{b}(E_{m\boldsymbol{k}}+E_{n\boldsymbol{k}})}{E_{m\boldsymbol{k}}-E_{n\boldsymbol{k}}}\xi_{nm}^{a}\xi_{mn}^{l}-i\sum_{s}\frac{E_{s\boldsymbol{k}}-E_{n\boldsymbol{k}}}{E_{m\boldsymbol{k}}-E_{n\boldsymbol{k}}}\xi_{mn}^{l}\mathcal{W}_{ns}^{a}\xi_{sm}^{b}-i\sum_{s}\frac{E_{m\boldsymbol{k}}-E_{s\boldsymbol{k}}}{E_{m\boldsymbol{k}}-E_{n\boldsymbol{k}}}\xi_{ns}^{b}\mathcal{W}_{sm}^{a}\xi_{mn}^{l}\Bigg\}.\label{nuTildeEa}
\end{align}

While (\ref{nuBarE}) and (\ref{nuTildeEb}) are generally gauge dependent, (\ref{nuTildeEa}) is only gauge dependent if there are degeneracies present in the unperturbed system. Very generally, there is a simplification that occurs when (\ref{nuBarE},\ref{nuTildeEb},\ref{nuTildeEa}) are summed to form the total response tensor (\ref{eq:beta_total}); the gauge-dependent terms appearing in (\ref{nuTildeEa}) cancel with terms appearing in (\ref{nuBarE}), and as a result the gauge-dependent terms appearing in the total $\beta_{\mathscr{M}}^{il}(\omega)$ do not explicitly depend on the energies $E_{n\boldsymbol{k}}$. In all we have
\begin{align}
& \beta_{\mathscr{M}}^{il}(\omega)=\alpha^{li}+\alpha_{\mathscr{M}}^{li}(\omega),\label{MtoE}
\end{align}
where we have separated out the dc-like terms, $\alpha^{li}=\alpha_{\text{G}}^{li}+\delta^{il}\alpha_{\text{CS}}$, as in (\ref{eq:betaP}), and defined 
\begin{align}
\alpha_{\mathscr{M}}^{li}(\omega)&\equiv\frac{e^{2}}{4c}\epsilon^{iab}\sum_{mn}f_{nm}\int_{\text{BZ}}\frac{d\boldsymbol{k}}{(2\pi)^{3}}\frac{\omega}{E_{m\boldsymbol{k}}-E_{n\boldsymbol{k}}-\hbar(\omega+i0^{+})}\Bigg\{2\frac{\partial_{b}(E_{m\boldsymbol{k}}+E_{n\boldsymbol{k}})}{E_{m\boldsymbol{k}}-E_{n\boldsymbol{k}}}\xi_{nm}^{a}\xi_{mn}^{l}+i\sum_{s}\frac{E_{s\boldsymbol{k}}-E_{m\boldsymbol{k}}}{E_{m\boldsymbol{k}}-E_{n\boldsymbol{k}}}\xi_{ns}^{a}\xi_{sm}^{b}\xi_{mn}^{l}\nonumber \\
&\quad\qquad\qquad\qquad\qquad\qquad\qquad\qquad+i\sum_{s}\frac{E_{n\boldsymbol{k}}-E_{s\boldsymbol{k}}}{E_{m\boldsymbol{k}}-E_{n\boldsymbol{k}}}\xi_{ns}^{b}\xi_{sm}^{a}\xi_{mn}^{l}-i\xi_{mn}^{l}\sum_{s}\Big(\mathcal{W}_{ns}^{a}\xi_{sm}^{b}+\xi_{ns}^{b}\mathcal{W}_{sm}^{a}\Big)\Bigg\}.\label{alphaM}
\end{align}
Like $\alpha_{\mathscr{P}}^{il}(\omega)$, $\alpha_{\mathscr{M}}^{li}(\omega)$ is \textit{entirely} a consequence of a dynamical contribution. The form of (\ref{alphaM}) is similar to that (\ref{alphaP}) found for $\alpha_{\mathscr{P}}^{il}(\omega)$, apart from a term related to an energy derivative. Also, like $\alpha^{il}_{\text{G}}$, $\alpha_{\mathscr{P}}^{il}(\omega)$ and $\alpha_{\mathscr{M}}^{li}(\omega)$ are ``cross-gap'' contributions; that is, they depend on \textit{both} initially occupied and unoccupied Bloch energy eigenstates, and their corresponding energies. Unlike $\alpha^{il}_{\text{G}}$, however, both $\alpha_{\mathscr{P}}^{il}(\omega)$ and $\alpha_{\mathscr{M}}^{li}(\omega)$ are gauge dependent.
\end{widetext}

A qualitative feature shared by the response tensors $\gamma^{ijl}(\omega)$, $\alpha_{\mathscr{P}}^{il}(\omega)$, $\alpha_{\mathscr{M}}^{li}(\omega)$, and $\chi^{ijl}_{\mathscr{Q}}(\omega)$ is that they are all gauge dependent. Moreover, the explicitly gauge-dependent terms within these tensors are of a similar form; the terms that involve the objects $\mathcal{W}_{nm}^a$ are all linear in $\mathcal{W}_{nm}^a$, and also involve the energies $E_{n\boldsymbol{k}}$ and the non-Abelian Berry connection $\xi_{nm}^b$. This is in contrast to what is found at the level of uniform and static Maxwell fields, where the only gauge dependence of such a tensor enters via the Chern-Simons contribution (\ref{alphaCS}) to the OMP tensor \cite{Mahon2020,Essin2010,Malashevich2010}. There the explicitly gauge-dependent term of $\alpha_{\text{CS}}$ involves the $\mathcal{W}^a$ alone and gives rise to a discrete ambiguity associated with the OMP tensor.

\section{Macroscopic charge and current densities}
\label{Section5}

We now construct expressions for the linear response of the macroscopic charge and current densities to the Maxwell fields, and as well identify the effective conductivity tensor $\sigma^{il}(\boldsymbol{q},\omega)$ to first order in $\boldsymbol{q}$. 

\subsection{The macroscopic current density}

Retaining only the contributions to the multipole moments that are linearly induced by the Maxwell fields and that are explicitly included in (\ref{eq:linear_response-2}), implementing them into the expressions (\ref{eq:Jandrho-1},\ref{eq:PandM-1}) to obtain the linear response of the current density and, following (\ref{eq:time_frequency}), writing this as 
\begin{align*}
\boldsymbol{J}^{(1)}(\boldsymbol{x},t)=\sum_{\omega}e^{-i\omega t}\boldsymbol{J}^{(1)}(\boldsymbol{x},\omega),
\end{align*}
we arrive at
\begin{align}
&J^{i(1)}(\boldsymbol{x},\omega)= \nonumber\\
&-i\omega\chi_{E}^{il}(\omega)E^{l}(\boldsymbol{x},\omega)-i\omega\gamma^{ijl}(\omega)F^{jl}(\boldsymbol{x},\omega)\nonumber \\
& -i\omega\big(\alpha^{il}+\alpha_{\mathscr{P}}^{il}(\omega)\big)B^{l}(\boldsymbol{x},\omega) +i\omega\chi_{\mathscr{Q}}^{ijl}(\omega)\frac{\partial E^{l}(\boldsymbol{x},\omega)}{\partial x^{j}}\nonumber \\
& +c\epsilon^{iab}\big(\alpha^{lb}+\alpha_{\mathscr{M}}^{lb}(\omega)\big)\frac{\partial E^{l}(\boldsymbol{x},\omega)}{\partial x^{a}},\label{inducedCurrent}
\end{align}
where $\alpha^{il}=\alpha^{il}_{\text{G}}+\delta^{il}\alpha_{\text{CS}}$. Of the response tensors appearing here, only $\chi_{E}^{il}(\omega)$ and $\alpha_{\text{G}}^{il}$ are gauge invariant. The rest, which are $\alpha_{\text{CS}}$, $\alpha_{\mathscr{P}}^{il}(\omega)$, $\alpha_{\mathscr{M}}^{il}(\omega)$, $\gamma^{ijl}(\omega)$, and $\chi_{\mathscr{Q}}^{ijl}(\omega)$, are all gauge dependent. Yet the linear response of the current density $\boldsymbol{J}^{(1)}(\boldsymbol{x},\omega)$ is in fact gauge invariant. To see this, first note that $\alpha_{\text{CS}}$ appears in (\ref{inducedCurrent}) in the form 
\begin{align*}
\alpha_{\text{CS}}\left(-i\omega B^{i}(\boldsymbol{x},\omega)+c\epsilon^{iab}\frac{\partial E^{b}(\boldsymbol{x},\omega)}{\partial x^{a}}\right)=0,
\end{align*}
vanishing via Faraday's law. So in considering the bulk response (\ref{inducedCurrent}) we can discard $\alpha_{\text{CS}}$, replacing $\alpha^{il}$ by $\alpha_{\text{G}}^{il}$. For the other gauge-dependent terms, we re-express each response tensor as a sum of a gauge-invariant contribution, denoted by a breve accent, and a gauge-dependent contribution. We then find
\begin{align}
&J^{i(1)}(\boldsymbol{x},\omega)=\nonumber\\
&-i\omega\chi_{E}^{il}(\omega)E^{l}(\boldsymbol{x},\omega)-i\omega\breve{\gamma}^{ijl}(\omega)F^{jl}(\boldsymbol{x},\omega)\nonumber \\
&-i\omega\big(\alpha_{\text{G}}^{il}+\breve{\alpha}_{\mathscr{P}}^{il}(\omega)\big)B^{l}(\boldsymbol{x},\omega)+i\omega\breve{\chi}_{\mathscr{Q}}^{ijl}(\omega)\frac{\partial E^{l}(\boldsymbol{x},\omega)}{\partial x^{j}}\nonumber \\
&+c\epsilon^{iab}\big(\alpha_{\text{G}}^{lb}+\breve{\alpha}_{\mathscr{M}}^{lb}(\omega)\big)\frac{\partial E^{l}(\boldsymbol{x},\omega)}{\partial x^{a}}\label{inducedCurrentDensity}
\end{align}
(see Appendix \ref{Appendix:InducedCurrent}); that is, the sum of the gauge-dependent contributions
vanishes. Thus the linear response of the current density is gauge invariant, as expected.

\subsection{The macroscopic charge density}

A similar analysis holds for the linear response of the charge density to the Maxwell fields, where from (\ref{eq:Jandrho-1}) we have
\begin{align*}
\varrho^{(1)}(\boldsymbol{x},t)=-\boldsymbol{\nabla}\boldsymbol{\cdot}\boldsymbol{P}^{(1)}(\boldsymbol{x},t).
\end{align*}
Again, retaining only the contributions to $\boldsymbol{P}(\boldsymbol{x},t)$ that are explicitly included in (\ref{eq:PandM-1}), those involving the electric dipole and quadrupole moments, and retaining only the contributions to the electric dipole and quadrupole moments that are linearly induced by the Maxwell fields and that are explicitly included in (\ref{eq:linear_response-2}), for the frequency components we have
\begin{align}
&\varrho^{(1)}(\boldsymbol{x},\omega)=\nonumber\\
&-\chi_{E}^{al}(\omega)\frac{\partial E^{l}(\boldsymbol{x},\omega)}{\partial x^{a}}-\gamma^{ajl}(\omega)\frac{\partial F^{jl}(\boldsymbol{x},\omega)}{\partial x^{a}}\nonumber \\
&-\big(\alpha^{al}+\alpha_{\mathscr{P}}^{al}(\omega)\big)\frac{\partial B^{l}(\boldsymbol{x},\omega)}{\partial x^{a}}+\chi_{\mathscr{Q}}^{ajl}(\omega)\frac{\partial^{2}E^{l}(\boldsymbol{x},\omega)}{\partial x^{a}\partial x^{j}}.\label{inducedCharge}
\end{align}
Again the Chern-Simons coefficient $\alpha_{\text{CS}}$ makes no contribution, since it appears in the form 
\begin{align*}
\alpha_{\text{CS}}\left(\frac{\partial B^{a}(\boldsymbol{x},\omega)}{\partial x^{a}}\right)=0,
\end{align*}
vanishing since the Maxwell magnetic field necessarily satisfies $\boldsymbol{\nabla}\boldsymbol{\cdot}\boldsymbol{B}(\boldsymbol{x},t)=0$. This is analogous to the scenario for $\boldsymbol{J}^{(1)}(\boldsymbol{x},\omega)$. As was the situation there, we expect (\ref{inducedCharge}) to be gauge invariant as a whole. Separating out the explicitly gauge-dependent terms as before, we find 
\begin{align}
&\varrho^{(1)}(\boldsymbol{x},\omega)=\nonumber\\
&-\chi_{E}^{al}(\omega)\frac{\partial E^{l}(\boldsymbol{x},\omega)}{\partial x^{a}}-\breve{\gamma}^{ajl}(\omega)\frac{\partial F^{jl}(\boldsymbol{x},\omega)}{\partial x^{a}}\nonumber \\
&-\big(\alpha_{\text{G}}^{al}+\breve{\alpha}_{\mathscr{P}}^{al}(\omega)\big)\frac{\partial B^{l}(\boldsymbol{x},\omega)}{\partial x^{a}}+\breve{\chi}_{\mathscr{Q}}^{ajl}(\omega)\frac{\partial^{2}E^{l}(\boldsymbol{x},\omega)}{\partial x^{a}\partial x^{j}}
\label{inducedChargeDensity}
\end{align}
(see Appendix \ref{Appendix:InducedCharge}), which is gauge invariant, as expected. 

We note that the expressions (\ref{inducedCurrentDensity},\ref{inducedChargeDensity}) satisfy continuity
\begin{align}
-i\omega\varrho^{(1)}(\boldsymbol{x},\omega)+\frac{\partial J^{i{(1)}}(\boldsymbol{x},\omega)}{\partial x^{i}}=0,
\end{align}
as also expected.

\subsection{The effective conductivity tensor}
\label{Section5c}

Finally, we can identify the linear dependence on $\boldsymbol{q}$ of the effective conductivity tensor $\sigma^{il}(\boldsymbol{q},\omega)$. Fourier transforming (\ref{eq:Jexpand}) to position space, we have
\begin{align*}
J^{i(1)}(\boldsymbol{x},\omega) & =\sigma^{il}(\omega)E^{l}(\boldsymbol{x},\omega)-i\sigma^{ilj}(\omega)\frac{\partial E^{l}(\boldsymbol{x},\omega)}{\partial x^{j}}+\ldots
\end{align*}
Comparing with (\ref{inducedCurrentDensity}) we can identify
\begin{align}
\sigma^{il}(\omega)=-i\omega\chi_{E}^{il}(\omega),\label{eq:silresult}
\end{align}
which agrees with the usual optical conductivity tensor found via the Kubo formula in the long wavelength limit \cite{Mahon2019}. Then, defining the dc limit of $\sigma^{ilj}(\omega)$ as
\begin{align}
\sigma_{\text{DC}}^{ilj}\equiv-ic\alpha_{\text{G}}^{ia}\epsilon^{ajl}+ic\epsilon^{ijb}\alpha_{\text{G}}^{lb} \label{eq:silkdc}
\end{align}
and implementing Faraday's law, we can identify
\begin{align}
\sigma^{ilj}(\omega) & =\sigma_{\text{DC}}^{ilj}+\omega\breve{\gamma}^{ijl}(\omega)-\omega\breve{\chi}_{\mathscr{Q}}^{ijl}(\omega)\nonumber\\
& \quad-ic\breve{\alpha}_{\mathscr{P}}^{ia}(\omega)\epsilon^{ajl}+ic\epsilon^{ijb}\breve{\alpha}_{\mathscr{M}}^{lb}(\omega);
\label{eq:silk} 
\end{align}
note $\breve{\gamma}^{ijl}(\omega)=\breve{\gamma}^{ilj}(\omega)$ and $\breve{\chi}_{\mathscr{\mathscr{Q}}}^{ijl}(\omega)=\breve{\chi}_{\mathscr{Q}}^{jil}(\omega)$. All of (\ref{eq:silresult},\ref{eq:silkdc},\ref{eq:silk}) are gauge invariant, as expected.

In the absence of time-reversal symmetry, the $\sigma^{il}(\omega)$ of (\ref{eq:silresult}) is nonsymmetric and can lead to the rotation of the plane of polarization of light as it propagates through the medium; this can be thought of as an ``internal'' Faraday effect, as illustrated by the discussion of the molecular crystal limit in the next section. The $\sigma^{ilj}(\omega)$ of (\ref{eq:silk}) is generally nonvanishing and nonsymmetric with respect to the exchange of any of its indices, even in the presence of time-reversal symmetry. But if that symmetry is present, then $\sigma^{ilj}_{\text{DC}}$ will vanish and the resulting $\sigma^{ilj}(\omega)$ describes what has been called \textit{natural} optical activity \cite{Souza2010}. In general the tensor $\sigma^{ilj}(\omega)$ can be evaluated at frequencies above the band gap, and thus can be used to describe both optical rotary dispersion and circular dichroism. Earlier work \cite{Souza2010} considered $\sigma^{ilj}(\omega)$ at frequencies below the band gap, where $E_{c\boldsymbol{k}}-E_{v\boldsymbol{k}}\neq\hbar\omega$ for all $c$, $v$, and $\boldsymbol{k}$; here ($c$) $v$ are the band indices labeling Bloch energy eigenstates of the unperturbed Hamiltonian that are initially (un)occupied. To compare our results with theirs, in our expression (\ref{eq:silk}) for $\sigma^{ilj}(\omega)$ we can take the $0^{+}$ limit immediately without introducing any divergences, and we follow them \cite{Souza2010} in adopting the notation ``$\doteq$'' to identify equalities that only formally hold in this limit. Introducing the shorthand $E_{cv\boldsymbol{k}}\equiv E_{c\boldsymbol{k}}-E_{v\boldsymbol{k}}$, and putting $\sigma^{ilj}(\omega)=\text{Re}[\sigma^{ilj}(\omega)]+i\text{Im}[\sigma^{ilj}(\omega)]$, we find 
\begin{align}
& \text{Re}[\sigma^{ilj}(\omega)]\doteq\nonumber \\
&e^{2}\sum_{cv}\int_{\text{BZ}}\frac{d\boldsymbol{k}}{(2\pi)^{3}}\Bigg\{2\frac{\hbar\omega}{E_{cv\boldsymbol{k}}^{2}-(\hbar\omega)^{2}}\text{Re}\Big[\mathcal{B}_{vc}^{ij}\xi_{cv}^{l}-\xi_{cv}^{i}\mathcal{B}_{vc}^{lj}\Big]\nonumber \\
&\qquad-\frac{\omega\big(3E_{cv\boldsymbol{k}}^{2}-(\hbar\omega)^{2}\big)}{\big(E_{cv\boldsymbol{k}}^{2}-(\hbar\omega)^{2}\big)^{2}}\partial_{j}(E_{c\boldsymbol{k}}+E_{v\boldsymbol{k}})\text{Im}\Big[\xi_{vc}^{i}\xi_{cv}^{l}\Big]\Bigg\}\label{realSigma}
\end{align}
and
\begin{align}
& \text{Im}[\sigma^{ilj}(\omega)]\doteq\nonumber \\
&2e^{2}\sum_{cv}\int_{\text{BZ}}\frac{d\boldsymbol{k}}{(2\pi)^{3}}\Bigg\{\frac{E_{cv\boldsymbol{k}}}{E_{cv\boldsymbol{k}}^{2}-(\hbar\omega)^{2}}\text{Im}\Big[\mathcal{B}_{vc}^{ij}\xi_{cv}^{l}+\xi_{cv}^{i}\mathcal{B}_{vc}^{lj}\Big]\nonumber \\
&\qquad+\frac{E_{cv\boldsymbol{k}}^{3}}{\big(E_{cv\boldsymbol{k}}^{2}-(\hbar\omega)^{2}\big)^{2}}\partial_{j}(E_{c\boldsymbol{k}}+E_{v\boldsymbol{k}})\text{Re}\Big[\xi_{vc}^{i}\xi_{cv}^{l}\Big]\Bigg\},\label{imaginarySigma}
\end{align}
where we have adopted the previously introduced \cite{Souza2010}
\begin{align}
\mathcal{B}_{nm}^{ab}&\equiv-\frac{i}{2\hbar}\partial_{a}(E_{n\boldsymbol{k}}+E_{m\boldsymbol{k}})\xi_{nm}^{b} \nonumber \\
&+\frac{1}{2\hbar}\sum_{s}\Big(E_{ns\boldsymbol{k}}\xi_{ns}^{a}\xi_{sm}^{b}+E_{sm\boldsymbol{k}}\xi_{ns}^{b}\xi_{sm}^{a}\Big)
\label{calB}
\end{align}
(see Appendix \ref{Appendix:Conductivity}). This is in agreement with the orbital electronic contribution to $\sigma^{ilj}(\omega)$ found by Malashevich and Souza \cite{Souza2010}, as expected. Notably the only nonvanishing contribution to $\sigma^{ilj}(\omega)$ in the $\omega\rightarrow0$ limit is due to $\sigma_{\text{DC}}^{ilj}$, which is purely imaginary, as $\alpha_{\text{G}}^{il}$ is real. Thus, in this limit, (\ref{realSigma}) is expected to vanish, which it does.

\section{The molecular crystal limit}
\label{Section6}

We now consider our response tensors $\chi_{E}^{il}(\omega)$, $\gamma^{ijl}(\omega)$, $\chi_{\mathscr{Q}}^{ijl}(\omega)$, $\alpha^{il}=\alpha_{\text{G}}^{il}+\delta^{il}\alpha_{\text{CS}}$, $\alpha_{\mathscr{P}}^{il}(\omega)$, and $\alpha_{\mathscr{M}}^{li}(\omega)$ in the molecular crystal limit. That is, we consider a periodic array of molecules where the orbitals associated with a molecule at a given lattice site share no common support with those of molecules associated with other lattice sites; again, we take the external electric and magnetic fields to which the molecules respond to be the macroscopic Maxwell fields, neglecting any local field corrections. We denote the response tensors in this limit by a circle accent.

We discussed the approach to this limit from the full crystalline expressions earlier \cite{Mahon2020}; in essence, this limit can be reached by taking the ELWFs (\ref{WF}) to be eigenfunctions of $H_{0}\big(\boldsymbol{x},\boldsymbol{\mathfrak{p}}(\boldsymbol{x})\big)$, in addition to the condition on the common support of these functions mentioned above \footnote{While we refer to this processing as a ``limit'', it is more accurately described as an imposed set of approximations on the ELWFs constructed via (\ref{WF}). Implementing these approximations in the above expressions is expected to lead to the reduction of such expressions to those derived for a periodic array of isolated atoms. It is \textit{not} the case that this set of approximations can simultaneously be valid for all crystalline systems for which ELWFs can be constructed in this manner. However for ``ordinary'' insulators that are adiabatically connected to the isolated atom limit, these approximations may hold.}. The former condition can be achieved by taking $E_{n\boldsymbol{k}}\rightarrow E_{n}$ and $U_{n\alpha}(\boldsymbol{k})\rightarrow\delta_{n\alpha}$, and, consequently,
\begin{align*}
& \xi_{cv}^{a}(\boldsymbol{k})=i\left(c\boldsymbol{k}|\partial_{a}v\boldsymbol{k}\right)\rightarrow x_{cv}^{a},
\end{align*}
where 
\begin{align}
x_{cv}^{a}\equiv\int W_{c\boldsymbol{0}}^{*}(\boldsymbol{x})x^{a}W_{v\boldsymbol{0}}(\boldsymbol{x})d\boldsymbol{x}.
\label{eq:position}
\end{align}
Again restricting ourselves to frequencies below the band gap, as in the second part of Section \ref{Section5c}, and implementing these substitutions, (\ref{chiE}) becomes 
\begin{align}
\mathring{\chi}_{E}^{il}(\omega) & \doteq\frac{e^{2}}{\Omega_{uc}}\sum_{vc}\Bigg(\frac{x_{vc}^{i}x_{cv}^{l}}{E_{cv}-\hbar\omega}+\frac{x_{vc}^{l}x_{cv}^{i}}{E_{cv}+\hbar\omega}\Bigg).\label{eq:PEmcl}
\end{align}
where $E_{n'n}\equiv E_{n'}-E_{n}$. In the presence of time-reversal symmetry this tensor is symmetric under the exchange of Cartesian components $i$ and $l$, but in general it is not, and we have only $\mathring{\chi}_{E}^{il}(\omega)\doteq\mathring{\chi}_{E}^{li}(-\omega) $. These results follow the pattern of the corresponding tensor for the more general crystalline system (see text surrounding Eq.~(\ref{chiE})). Note that even were it the only response tensor present, an asymmetric $\chi_{E}^{il}(\omega)$ would be sufficient to lead to the rotation of the polarization of light as it propagates through a medium, as can be easily confirmed. In this molecular crystal limit it is easy to give an example of how this might arise. Suppose, for example, that the breaking of time-reversal necessary for the asymmetric $\mathring{\chi}_{E}^{il}(\omega)$ occurs because each molecule -- or, simpler, atom -- is subject to a dc magnetic field that is incorporated in the unperturbed atomic Hamiltonian. Then, if light is propagating in the direction of the dc magnetic field, then the rotation of its plane of polarization that results is just the Faraday effect, which is well known in atomic systems and indeed has a variety of applications \cite{Budker}. Next, (\ref{gamma}) simplifies to 
\begin{align}
\mathring{\gamma}^{ijl}(\omega) & \doteq\frac{e^{2}}{2\Omega_{uc}}\sum_{vcn}\Bigg(\frac{x_{vc}^{i}x_{cn}^{j}x_{nv}^{l}}{E_{cv}-\hbar\omega}+\frac{x_{vn}^{j}x_{nc}^{l}x_{cv}^{i}}{E_{cv}+\hbar\omega}\Bigg),\label{eq:PFmcl}
\end{align}
and (\ref{chiQ}) to 
\begin{align}
\mathring{\chi}_{\mathscr{Q}}^{ijl}(\omega)\doteq\frac{e^{2}}{2\Omega_{uc}}\sum_{vcn}\Bigg(\frac{x_{vn}^{i}x_{nc}^{j}x_{cv}^{l}}{E_{cv}-\hbar\omega}+\frac{x_{vc}^{l}x_{cn}^{i}x_{nv}^{j}}{E_{cv}+\hbar\omega}\Bigg).\label{eq:QEmcl}
\end{align}
Recall from previous work \cite{Mahon2020} 
\begin{align*}
\mathring{\alpha}^{il} & =\frac{e^{2}}{2mc\Omega_{uc}}\epsilon^{lab}\sum_{vcn}\Bigg(\frac{x_{vc}^{i}x_{cn}^{a}\mathfrak{p}_{nv}^{b}}{E_{cv}}+\frac{x_{vn}^{a}\mathfrak{p}_{nc}^{b}x_{cv}^{i}}{E_{cv}}\Bigg),
\end{align*}
where, in this limit, 
\begin{align}
\boldsymbol{\mathfrak{p}}_{n'n}\equiv\frac{im}{\hbar}E_{n'n}\boldsymbol{x}_{n'n}.
\label{eq:momentum}
\end{align}
Further, (\ref{alphaP}) becomes 
\begin{align*}
& \mathring{\alpha}_{\mathscr{P}}^{il}(\omega)\doteq\\
&\frac{e^{2}\hbar\omega}{2mc\Omega_{uc}}\epsilon^{lab}\sum_{vcn}\Bigg(\frac{x_{vc}^{i}x_{cn}^{a}\mathfrak{p}_{nv}^{b}}{E_{cv}(E_{cv}-\hbar\omega)}-\frac{x_{vn}^{a}\mathfrak{p}_{nc}^{b}x_{cv}^{i}}{E_{cv}(E_{cv}+\hbar\omega)}\Bigg).
\end{align*}
Then, combining this with the dc-like contribution, the full response of the polarization to the magnetic field is given by
\begin{align}
\mathring{\beta}_{\mathscr{P}}^{il}(\omega)&= \mathring{\alpha}^{il}+\mathring{\alpha}_{\mathscr{P}}^{il}(\omega)\nonumber\\
&\doteq\frac{e^{2}}{2mc\Omega_{uc}}\epsilon^{lab}\sum_{vcn}\Bigg(\frac{x_{vc}^{i}x_{cn}^{a}\mathfrak{p}_{nv}^{b}}{E_{cv}-\hbar\omega}+\frac{x_{vn}^{a}\mathfrak{p}_{nc}^{b}x_{cv}^{i}}{E_{cv}+\hbar\omega}\Bigg).\label{eq:PBmcl} 
\end{align}
Finally, (\ref{alphaM}) simplifies to 
\begin{align*}
& \mathring{\alpha}_{\mathscr{M}}^{li}(\omega)\doteq\\
&\frac{e^{2}\hbar\omega}{2mc\Omega_{uc}}\epsilon^{iab}\sum_{vcn}\Bigg(\frac{x_{vn}^{a}\mathfrak{p}_{nc}^{b}x_{cv}^{l}}{E_{cv}(E_{cv}-\hbar\omega)}-\frac{x_{cn}^{a}\mathfrak{p}_{nv}^{b}x_{vc}^{l}}{E_{cv}(E_{cv}+\hbar\omega)}\Bigg).
\end{align*}
Combining this with the dc-like contribution, the full response of the magnetization to the electric field is given by
\begin{align}
\mathring{\beta}_{\mathscr{M}}^{il}(\omega)&=\mathring{\alpha}^{li}+\mathring{\alpha}_{\mathscr{M}}^{li}(\omega) \nonumber\\
&\doteq\frac{e^{2}}{2mc\Omega_{uc}}\epsilon^{iab}\sum_{vcn}\Bigg(\frac{x_{vn}^{a}\mathfrak{p}_{nc}^{b}x_{cv}^{l}}{E_{cv}-\hbar\omega}+\frac{x_{vc}^{l}x_{cn}^{a}\mathfrak{p}_{nv}^{b}}{E_{cv}+\hbar\omega}\Bigg).\label{eq:MEmcl} 
\end{align}

Physically one expects that an equivalent way to derive these expressions would be to solve for the linearly induced moments of the individual molecules; since local field corrections are being neglected, the fields to which they respond are the Maxwell fields, and the limiting response tensors above should be equal to the appropriate molecular response tensors multiplied by the number of molecules per unit volume, here equal to $\Omega_{uc}^{-1}$. The molecular calculations can be made with the usual multipole moment Hamiltonian \cite{PZW,Healybook}, which including the moments relevant here can be written as
\begin{align*}
\hat{H}_{\text{mol}}(t)&=\hat{H}_{\text{mol}}^{0}-\hat{\mu}^{i}E^{i}(t)-\hat{q}^{ij}F^{ij}(t)\\
&-\hat{\nu}_{P}^{i}B^{i}(t)-\frac{1}{2}\hat{\nu}_{D}^{i}B^{i}(t),
\end{align*}
where $\hat{H}_{\text{mol}}^{0}$ is the Hamiltonian in the absence of any Maxwell fields; $E^{i}(t)$, $F^{ij}(t)$, and $B^{i}(t)$ are the Cartesian components of the electric field, its symmeterized derivative, and the magnetic field evaluated at the position of the molecule; and $\hat{\mu}^{i}$, $\hat{q}^{ij}$, $\hat{\nu}_{P}^{i}$, and $\hat{\nu}_{D}^{i}$ are the indicated components of the operators for the electric dipole and quadrupole moments, and the paramagnetic and diamagnetic dipole moments of the molecule. The diagmagnetic dipole moment is neglected here since it is not involved in optical activity, but the matrix elements of the other moments can be written in terms of the ``position'' and ``momentum'' matrix elements (\ref{eq:position},\ref{eq:momentum}) involving the $\left\{W_{v\boldsymbol{0}}(\boldsymbol{x})\right\}$ and the $\left\{W_{c\boldsymbol{0}}(\boldsymbol{x})\right\}$, now identified with the filled and empty orbitals of a molecule fixed at the origin.

The result is that (\ref{eq:PEmcl},\ref{eq:PFmcl},\ref{eq:QEmcl},\ref{eq:PBmcl},\ref{eq:MEmcl}) are indeed the appropriate molecular response tensors divided by $\Omega_{uc}$. The molecular calculation also clarifies certain symmetries in the expressions in the molecular crystal limit. For example, in this case one can immediately identify
\begin{align*}
\mathring{\chi}_{\mathscr{Q}}^{ijl}(\omega)&\doteq\mathring{\gamma}^{lij}(-\omega),
\end{align*}
and the equivalent expressions
\begin{align*}
\mathring{\beta}_{\mathscr{P}}^{il}(\omega)&\doteq\mathring{\beta}_{\mathscr{M}}^{li}(-\omega),\\
\mathring{\alpha}_{\mathscr{P}}^{il}(\omega)&\doteq\mathring{\alpha}_{\mathscr{M}}^{il}(-\omega).
\end{align*}
The first of these holds because the response calculations leading to both quantities involve the different-time commutator of the electric dipole and the electric quadrupole moment operators, while the second holds because the response calculations leading to both involve the different-time commutator of the electric dipole and the paramagnetic dipole moment operators. These symmetries no longer hold in the full crystal calculation, where the site multipole moments are \textit{not} the result of the expectation values of site operators, but rather are evaluated in terms of the single-particle Green function.

\section{Conclusion}
\label{Section7}

We have presented a theory for the effective conductivity tensor $\sigma^{il}(\boldsymbol{q},\omega)$ of a class of insulating crystalline solids at zero temperature. In retaining terms that are at most linear in $\boldsymbol{q}$, we extract tensors $\sigma^{il}(\omega)$ and $\sigma^{ilj}(\omega)$ that describe phenomena involving the rotation of the plane of polarization of light as it propagates through a medium; the former contributes through its antisymmetric part only when time-reversal symmetry is broken in the unperturbed system, and can be considered as describing an ``internal'' Faraday effect, while the latter contributes more generally and describes optical activity. Although we have restricted ourselves to the independent particle approximation, and have neglected spin effects and the motion of ion cores, within these approximations our expression for $\sigma^{ilj}(\omega)$ describes both optical rotary dispersion and circular dichroism.

Our approach is based on introducing microscopic polarization and magnetization fields, from which the charge and current density expectation values can be found. The corresponding macroscopic fields of elementary electrodynamics can then be defined as the spatial averages of those microscopic fields; the ``free'' macroscopic charge and current densities that can generally arise vanish in the linear response of the class of insulators we consider. With the use of a set of Wannier functions, we associate portions of these microscopic fields with each lattice site, thereby introducing site polarization and magnetization fields from which site multipole moments are extracted. 

We then construct macroscopic multipole moments from these site multipole moments, and from their linear response to the electromagnetic field we identify the tensors describing the response of the electric dipole moment per unit volume $\mathscr{P}^i(\boldsymbol{x},t)$ to the electric field $E^l(\boldsymbol{x},t)$, to the symmetrized derivative of the electric field $F^{jl}(\boldsymbol{x},t)$, and to the magnetic field $B^l(\boldsymbol{x},t)$; the response of the electric quadrupole moment per unit volume $\mathscr{Q}^{ij}(\boldsymbol{x},t)$ to $E^l(\boldsymbol{x},t)$; and the response of the magnetic dipole moment per unit volume $\mathscr{M}^i(\boldsymbol{x},t)$ to $E^l(\boldsymbol{x},t)$. From these tensors we construct $\sigma^{il}(\omega)$ and $\sigma^{ilj}(\omega)$. Due to its focus on identifying site quantities, our strategy allows for an easy comparison with results in the ``molecular crystal limit'', where the electrons associated with a molecule at one lattice site cannot move to another site. But it certainly does not require that idealization.

In the limit of uniform and static electric and magnetic fields we recover the magnetoelectric effect described earlier by others \cite{Essin2010,Malashevich2010} and us \cite{Mahon2020}, the latter calculation using the approach implemented here. There the first-order modifications of both $P^i$ due to $B^l$ and of $M^l$ due to $E^i$ are described by the orbital magnetoelectric polarizability (OMP) tensor $\alpha^{il}$, which is nonvanishing only if both time-reversal and spatial inversion symmetry are broken in the unperturbed system. At finite frequencies the previously identified contributions to the OMP tensor, $\alpha^{il}_{\text{G}}$ and $\delta^{il}\alpha_{\text{CS}}$, remain as contributions to the response of $\mathscr{P}^i(\boldsymbol{x},\omega)$ to $B^l(\boldsymbol{x},\omega)$ and of $\mathscr{M}^l(\boldsymbol{x},\omega)$ to $E^i(\boldsymbol{x},\omega)$. However, additional explicitly frequency-dependent contributions, $\alpha^{il}_{\mathscr{P}}(\omega)$ and $\alpha^{il}_{\mathscr{M}}(\omega)$, to the total response tensors emerge, generally resulting in the tensors describing the linear response of $\mathscr{P}^i(\boldsymbol{x},\omega)$ to $B^l(\boldsymbol{x},\omega)$ and of $\mathscr{M}^l(\boldsymbol{x},\omega)$ to $E^i(\boldsymbol{x},\omega)$ to differ. These additional contributions are classified as ``cross-gap'' contributions, like $\alpha^{il}_{\text{G}}$, but are gauge dependent. Thus, the net cross-gap contributions would be given by $\alpha^{il}_{\text{G}}+\alpha_{\mathscr{P}}^{il}(\omega)$ and $\alpha^{il}_{\text{G}}+\alpha_{\mathscr{M}}^{il}(\omega)$, respectively. The terms $\alpha_{\mathscr{P}}^{il}(\omega)$ and $\alpha_{\mathscr{M}}^{il}(\omega)$ that arise and differentiate the responses result from contributions that we identify as ``dynamical.'' Furthermore, as the finite frequency generalization of the ``compositional'' contributions to the response tensors is trivial, and because $\alpha^{il}_{\mathscr{P}}(\omega)$ and $\alpha^{il}_{\mathscr{M}}(\omega)$ are manifestly ``cross-gap'' contributions, the Chern-Simons contribution to the finite frequency response tensors remains unchanged; that is, the finite frequency generalization of the Chern-Simons contribution to these response tensors is identical to that in the limit of uniform and static Maxwell fields.

In the molecular crystal limit the Chern-Simons contribution, which does not contribute to the bulk macroscopic charge and current densities that are linearly induced by the Maxwell fields, becomes gauge invariant \footnote{The gauge-dependent term of $\alpha_{\text{CS}}$ can be used to identify ``ordinary'' and ``$\mathbb{Z}_2$-odd'' insulators \cite{Vanderbilt2009}. As this term vanishes under this set of approximations, they cannot simultaneously be satisfied for $\mathbb{Z}_2$-odd insulators.}. As well, in that limit the response tensor characterizing the finite frequency linear response of $\mathscr{P}^i(\boldsymbol{x},\omega)$ to $B^l(\boldsymbol{x},\omega)$, and the response tensor characterizing that of $\mathscr{M}^l(\boldsymbol{x},\omega)$ to $E^i(\boldsymbol{x},\omega)$, are related; this relation does not hold in general beyond the molecular crystal limit. Similarly, the relations between the tensors describing the linear response of $\mathscr{P}^i(\boldsymbol{x},\omega)$ to $F^{jl}(\boldsymbol{x},\omega)$ and of $\mathscr{Q}^{ij}(\boldsymbol{x},\omega)$ to $E^l(\boldsymbol{x},\omega)$ that hold in the molecular crystal limit do not hold generally. This is because in the molecular crystal limit the site multipole moments can be associated with expectation values of associated operators familiar from molecular physics, whereas for a crystal in which charges can move more freely a Green function approach was used to define them.

Generally, these macroscopic multipole moments were introduced with the use of Wannier functions associated with each lattice site, and thus $\mathscr{P}^i(\boldsymbol{x},t)$, $\mathscr{Q}^{ij}(\boldsymbol{x},t)$, and $\mathscr{M}^i(\boldsymbol{x},t)$ are ``gauge dependent'' in the sense that they depend on the choice of these Wannier functions. A natural choice, of course, would be a set of ELWFs. However, we showed that whatever choice is made the expressions for the linear response of the macroscopic charge and current densities to the Maxwell fields are gauge invariant. Thus our expression for $\sigma^{ilj}(\omega)$, as well as that for $\sigma^{il}(\omega)$, can be evaluated without any calculation -- or any thought -- of the Wannier functions than underpin our approach. At frequencies below the band gap we found agreement with earlier work restricted to that frequency range \cite{Souza2010}.

Yet, while they do not appear explicitly in the final expression for $\sigma^{il}(\omega)$ or $\sigma^{ilj}(\omega)$, it is the site multipole moments that can be introduced with the aid of these Wannier functions, and the microscopic polarization and magnetization fields on which the approach is based, that make possible the natural connection and comparison with the molecular crystal limit. This should lead to an understanding of which features of the optical activity of any material of interest can be associated with physics beyond that limit. As well, the use of such site quantities in our approach offers the possibility of considering the optical response of a finite system, where simply identifying the bulk tensors $\sigma^{il}(\omega)$ and $\sigma^{ilj}(\omega)$ is not sufficient, and will lead to the description of other linear and nonlinear optical response features that depend on the variation of the electromagnetic field throughout a finite crystal. We plan to turn to these generalizations in future publications.

\section{Acknowledgments}
This work was supported by the Natural Sciences and Engineering Research Council of Canada (NSERC). P.T.M. acknowledges a PGS-D scholarship from NSERC.

\section{Appendices}
\appendix

\section{Formal relator expansions}
\label{Appendix:Relator}
The ``relators'' allow us to obtain the microscopic polarization and magnetization fields from the charge and current density expectation values. They also arise in the way the Maxwell fields enter the dynamical equations governing such quantities. Thus, an expansion of the relators is relevant for the identification of the electric and magnetic moments \textit{and} in expanding the equations of motion of quantities associated with the electron field in terms of powers of the Maxwell fields and their derivatives. As a consequence, the expansion parameter $u$ appearing in the relator expansions can be used to identify which perturbative modifications to the various site multipole moments due to a particular Maxwell field, or derivative of that field, appear at the same ``order.'' We now show this.

The expansions of $\Omega^j_{\boldsymbol{y}}(\boldsymbol{x},t)$
and $\Omega_{\boldsymbol{y}}^{0}(\boldsymbol{x},t)$, (\ref{eq:Omega},\ref{eq:omega0}), derived previously can be more easily derived using a formal expansion of the ``relators'', $s^{i}(\boldsymbol{w};\boldsymbol{x},\boldsymbol{y})$ and $\alpha^{ij}(\boldsymbol{w};\boldsymbol{x},\boldsymbol{y})$, about $u=0$. We begin with the definition, under the choice of a straight-line path; see Ref.~\cite{Mahon2019}, where we find 
\begin{align}
s^{i}(\boldsymbol{w};\boldsymbol{x},\boldsymbol{y}) & =\int_{0}^{1}(x^{i}-y^{i})\delta(\boldsymbol{w}-\boldsymbol{y}-u(\boldsymbol{x}-\boldsymbol{y}))du,\nonumber \\
\alpha^{ij}(\boldsymbol{w};\boldsymbol{x},\boldsymbol{y}) & =\epsilon^{iaj}\int_{0}^{1}(x^{a}-y^{a})\delta(\boldsymbol{w}-\boldsymbol{y}-u(\boldsymbol{x}-\boldsymbol{y}))udu.\label{eq:straight-line_use}
\end{align}
Recall we have previously defined
\begin{align}
& \Omega_{\boldsymbol{y}}^{j}(\boldsymbol{x},t)\equiv\int\alpha^{lj}(\boldsymbol{w};\boldsymbol{x},\boldsymbol{y})B^{l}(\boldsymbol{w},t)d\boldsymbol{w},\label{eq:OmegaVec}\\
& \Omega_{\boldsymbol{y}}^{0}(\boldsymbol{x},t)\equiv\int s^{i}(\boldsymbol{w};\boldsymbol{x},\boldsymbol{y})E^{i}(\boldsymbol{w},t)d\boldsymbol{w}, \label{eq:Omega_0} 
\end{align}
and found for nearly uniform Maxwell fields
\begin{align}
\Omega^a_{\boldsymbol{y}}(\boldsymbol{x},\omega)&\simeq\frac{\epsilon^{abd}}{2}B^b(\boldsymbol{y},\omega)\big(x^d-y^d\big)+\ldots,\label{omegaVec}\\
\Omega_{\boldsymbol{y}}^{0}(\boldsymbol{x},\omega)&\simeq\left(x^a-y^a\right)E^a(\boldsymbol{y},\omega)\nonumber\\
&+\frac{1}{2}(x^{a}-y^{a})(x^{b}-y^{b})F^{ab}(\boldsymbol{y},\omega)+\ldots\label{omega0Appendix},
\end{align}
which we have implemented in this work. We now find these approximate expressions in a different way. We write the first of (\ref{eq:straight-line_use}) as
\begin{align}
s^{i}(\boldsymbol{w};\boldsymbol{x},\boldsymbol{y})&\simeq\int_{0}^{1}(x^{i}-y^{i})\delta(\boldsymbol{w}-\boldsymbol{y})du \nonumber\\
&+\int_{0}^{1}(x^{i}-y^{i})u\left[\frac{\partial\delta(\boldsymbol{w}-\boldsymbol{y}-u(\boldsymbol{x}-\boldsymbol{y}))}{\partial u}\right]_{u=0}du \nonumber\\
&+\ldots \nonumber\\ 
&=(x^{i}-y^{i})\delta(\boldsymbol{w}-\boldsymbol{y}) \nonumber\\
&-\frac{1}{2}(x^{i}-y^{i})(x^{j}-y^{j})\frac{\partial\delta(\boldsymbol{w}-\boldsymbol{y})}{\partial w^{j}}+\ldots
\label{sExpansion}
\end{align}
Used in (\ref{eq:Omega_0}), following a partial
integration with respect to $\boldsymbol{w}$, immediately gives
(\ref{omega0Appendix}). Notice that the first term of (\ref{omega0Appendix}) 
originates from the $\mathscr{O}(u^0)$ term of the $s^i$-relator expansion (\ref{sExpansion}), and 
the second term from the $\mathscr{O}(u)$ term.

Similarly, we expand the second of (\ref{eq:straight-line_use}) to the same order, $\mathscr{O}(u)$, and find 
\begin{align}
\alpha^{ij}(\boldsymbol{w};\boldsymbol{x},\boldsymbol{y})&\simeq\epsilon^{iaj}\int_{0}^{1}(x^{a}-y^{a})\delta(\boldsymbol{w}-\boldsymbol{y})udu+\ldots \nonumber\\
& =\frac{1}{2}\epsilon^{iaj}(x^{a}-y^{a})\delta(\boldsymbol{w}-\boldsymbol{y})+\ldots,
\label{alphaExpansion}
\end{align}
Using this in (\ref{eq:OmegaVec}) we immediately arrive at (\ref{omegaVec}). Then the explicitly retained term of (\ref{omegaVec}) originates from an $\mathscr{O}(u)$ term of the $\alpha^{ij}$-relator expansion. Thus, (\ref{omegaVec}) and the second term of (\ref{omega0Appendix}) appear at the same order of the expansion parameter $u$. This is consistent with Faraday's law as the spatial derivatives of $\boldsymbol{E}(\boldsymbol{x},\omega)$ are related to frequency factors times $\boldsymbol{B}(\boldsymbol{x},\omega)$. Thus such terms appear at the same ``order'' with respect to the Maxwell fields and their derivatives kept in an expansion. It appears that the expansion parameter $u$ captures this information. 

Now the site electric and magnetic multipole moments can be found from the ``site'' polarization and magnetization fields, respectively, using these same relator expansions. The site electric dipole moment (\ref{eq:ED}) originates from the $\mathscr{O}(u^0)$ term of the $s^i$-relator expansion (\ref{sExpansion}), while the site electric quadrupole moment (\ref{eq:EQ}) originates from the $\mathscr{O}(u)$ term of the $s^i$-relator expansion. The site magnetic dipole moment (\ref{eq:MD}) originates from the $\mathscr{O}(u)$ term of the $\alpha^{ij}$-relator expansion (\ref{alphaExpansion}). However, it is \textit{not only} via the expansion of those relators that relate the microscopic charge and current densities to the microscopic polarization and magnetization fields that the expansion parameter $u$ enters. When finding the linear response of the single-particle density matrix, (\ref{EDMe},\ref{EDMf},\ref{EDMb2}), the quantities (\ref{omegaVec},\ref{omega0Appendix}) are used. 
Thus, modifications to the site electric dipole moment appear \textit{at least} at order $\mathscr{O}(u^0)$, but not all modifications to this quantity appear at this order; for instance, (\ref{muE}) appears at $\mathscr{O}(u^0)$, while (\ref{muBb}) and (\ref{muBa}) appear at $\mathscr{O}(u)$. Furthermore, modifications to the site electric quadrupole and the site magnetic dipole moments appear \textit{at least} at order $\mathscr{O}(u)$; for example, (\ref{chiQ}) and (\ref{nuBarE})-(\ref{nuTildeEa}) appear at $\mathscr{O}(u)$. In this work, we only consider the contributions to the linear response of a site quantity appearing \textit{at most} at $\mathscr{O}(u)$; we neglect higher-order modifications, such as those leading to the magnetic susceptibility, which would appear at $\mathscr{O}(u^2)$, those related to spatial derivatives of the magnetic field, or those related to the linear response of higher-order multipole moments.

\section{Microscopic and macroscopic fields}
\label{Appendix:MacroFields}

In this Appendix we describe approaches to constructing macroscopic fields from the microscopic fields appearing in (\ref{eq:micro}).

One approach ($I$) often adopted to treat infinite crystals is to start from the Fourier transform to wavevector space of all the quantities of interest. For the current density, for example, we would have
\begin{align*}
\boldsymbol{j}(\boldsymbol{q},t)\equiv\int e^{-i\boldsymbol{q}\boldsymbol{\cdot}\boldsymbol{x}}\boldsymbol{j}(\boldsymbol{x},t)d\boldsymbol{x},
\end{align*}
etc. If $\boldsymbol{j}(\boldsymbol{q},t)$ is nonzero only for a single, small $\boldsymbol{q}$, then the variation in the current density is sinusoidal. If one wants to consider less trivial variations, then one needs to treat a range of $\boldsymbol{q}$s. To do this one can introduce a macroscopic field associated with each microscopic field -- e.g., $\boldsymbol{J}(\boldsymbol{q},t)$ with $\boldsymbol{j}(\boldsymbol{q},t)$ -- by setting $\boldsymbol{J}(\boldsymbol{q},t)=\boldsymbol{j}(\boldsymbol{q},t)$ for some restricted region of reciprocal space near the origin -- say, $\left|\boldsymbol{q}\right|\leq\Delta^{-1}$, where $\Delta$ is a length satisfying $\Delta\gg a$, with $a$ on the order of a lattice constant -- and $\boldsymbol{J}(\boldsymbol{q},t)=\boldsymbol{0}$ for other $\boldsymbol{q}$ in reciprocal space. Also choosing $\Delta\ll\lambda$, where $\lambda$ characterizes a typical range of variation of the fields that one is trying to capture, the macroscopic fields can describe excitations in the crystal characterized by typical length scales much larger than the lattice constant.

Another approach ($II$) with the same goal starts in position space rather than reciprocal space, and introduces a smooth weighting function $\mathsf{w}(\boldsymbol{x})$ to extract a macroscopic field $L(\boldsymbol{x})$ from the associated microscopic field $l(\boldsymbol{x})$ \cite{Jackson}, 
\begin{align}
L(\boldsymbol{x})\equiv\int\mathsf{w}(\boldsymbol{x}-\boldsymbol{x}')l(\boldsymbol{x}')d\boldsymbol{x}',\label{eq:averaging}
\end{align}
identifying the macroscopic field at a point $\boldsymbol{x}$ with the average of the associated microscopic field in the neighborhood of $\boldsymbol{x}$. We take $\mathsf{w}(\boldsymbol{x})$ to be a smooth positive function, peaking at $\boldsymbol{x}=\boldsymbol{0}$ and spherically symmetric about that point, dropping off continuously as $\left|\boldsymbol{x}\right|\rightarrow\infty$ with a characteristic length scale $\Delta$ satisfying the conditions given above, 
\begin{align}
a\ll\Delta\ll\lambda,\label{eq:inequalities}
\end{align}
and with an integral over all space equal to unity. A typical example would be a Gaussian function, $\mathsf{w}(\boldsymbol{x})=\mathsf{w}_{II}(\boldsymbol{x})$, where 
\begin{align*}
\mathsf{w}_{II}(\boldsymbol{x})=\frac{e^{-\left|\boldsymbol{x}\right|^{2}/\Delta^{2}}}{\Delta^{3}\pi^{3/2}}.
\end{align*}

The two approaches can be formally related, of course, because from (\ref{eq:averaging}) we have 
\begin{align*}
L(\boldsymbol{q})=\mathsf{w}(\boldsymbol{q})l(\boldsymbol{q}),
\end{align*}
and by formally setting $\mathsf{w}_{I}(\boldsymbol{q})=\theta(\Delta^{-1}-\left|\boldsymbol{q}\right|)$, where $\theta(q)$ is the Heavyside step function, we recover the first approach. It has the advantage that constructing a macroscopic field from its associated microscopic field is a \textit{projection} in wavevector space; thus, choosing $\mathsf{w}(\boldsymbol{x})=\mathsf{w}_{I}(\boldsymbol{x})$, if the operation (\ref{eq:averaging}) is repeated there is no additional change. On the other hand, the $\mathsf{w}_{I}(\boldsymbol{x})$ that results 
\begin{align*}
\mathsf{w}_{I}(\boldsymbol{x})=\frac{1}{2\pi^{2}}\left(\frac{1}{\left|\boldsymbol{x}\right|^{3}}\sin\left(\frac{\left|\boldsymbol{x}\right|}{\Delta}\right)-\frac{\Delta}{\left|\boldsymbol{x}\right|^{2}}\cos\left(\frac{\left|\boldsymbol{x}\right|}{\Delta}\right)\right),
\end{align*}
extends far beyond $\left|\boldsymbol{x}\right|=\Delta$, and as well takes on negative values. Indeed, any $\mathsf{w}(\boldsymbol{q})$ which, like $\mathsf{w}_{I}(\boldsymbol{q})$, has a vanishing second derivative in some direction $\boldsymbol{\hat{q}}$ about \textbf{$\boldsymbol{q}=\boldsymbol{0}$} will lead to a $\mathsf{w}(\boldsymbol{x})$ which \textit{must} take on negative values, since it has a vanishing second moment. Thus the second approach, where one begins with a smooth and well-behaved averaging function in position space (and note $\mathsf{w}_{II}(\boldsymbol{q})=\text{exp}\big(-\left|\boldsymbol{q}\right|^{2}/(4\Delta^{2})\big)$), seems a better choice if one wants to understand the averaging physically, and with it one can envision a treatment of finite media and interfaces. In this paper we only concern ourselves with nominally infinite crystals, so the two approaches lead to essentially the same results; we indicate the small differences below, but most of what we say would apply to either.

We adopt the semiclassical approximation, where the electromagnetic field is treated classically, and in the Maxwell equations for the microscopic electric and magnetic fields, $\boldsymbol{e}(\boldsymbol{x},t)$ and $\boldsymbol{b}(\boldsymbol{x},t)$, we take $\rho(\boldsymbol{x},t)$ and $\boldsymbol{j}(\boldsymbol{x},t)$ (\ref{eq:micro}) as the microscopic charge-current density of the crystal. Using the averaging procedure (\ref{eq:averaging}) to identify the macroscopic fields from their microscopic counterparts, we immediately find that those macroscopic fields satisfy the macroscopic Maxwell equations in the form 
\begin{align}
&\nabla\boldsymbol{\cdot}\boldsymbol{D}(\boldsymbol{x},t)=4\pi\varrho_{F}(\boldsymbol{x},t),\nonumber\\ &c\nabla\cross\boldsymbol{H}(\boldsymbol{x},t)=4\pi\boldsymbol{J}_{F}(\boldsymbol{x},t)+\frac{\partial}{\partial t}\boldsymbol{D}(\boldsymbol{x},t),\nonumber \\
&\nabla\boldsymbol{\cdot}\boldsymbol{B}(\boldsymbol{x},t)=0,\nonumber \\
&c\nabla\cross\boldsymbol{E}(\boldsymbol{x},t)+\frac{\partial}{\partial t}\boldsymbol{B}(\boldsymbol{x},t)=0,\label{eq:macroMaxwell} 
\end{align}
where $\boldsymbol{D}(\boldsymbol{x},t)=\boldsymbol{E}(\boldsymbol{x},t)+4\pi\boldsymbol{P}(\boldsymbol{x},t)$,
$\boldsymbol{H}(\boldsymbol{x},t)=\boldsymbol{B}(\boldsymbol{x},t)-4\pi\boldsymbol{M}(\boldsymbol{x},t)$,
and 
\begin{align}
\varrho_{F}(\boldsymbol{x},t)&\equiv\int\mathsf{w}(\boldsymbol{x}-\boldsymbol{x}')\rho_{F}(\boldsymbol{x}',t)d\boldsymbol{x}',\nonumber\\
\boldsymbol{E}(\boldsymbol{x},t)&\equiv\int\mathsf{w}(\boldsymbol{x}-\boldsymbol{x}')\boldsymbol{e}(\boldsymbol{x}',t)d\boldsymbol{x}',\nonumber \\
\boldsymbol{P}(\boldsymbol{x},t)&\equiv\int\mathsf{w}(\boldsymbol{x}-\boldsymbol{x}')\boldsymbol{p}(\boldsymbol{x}',t)d\boldsymbol{x}',\label{eq:averaged} 
\end{align}
etc. As mentioned in the text, we refer to the macroscopic fields
$\boldsymbol{E}(\boldsymbol{x},t)$ and $\boldsymbol{B}(\boldsymbol{x},t)$
as the ``Maxwell fields.''

Using the expansions (\ref{eq:pmexpand}) in the expression (\ref{eq:total_pandm})
for the total $\boldsymbol{p}(\boldsymbol{x},t)$ and $\boldsymbol{m}(\boldsymbol{x},t)$,
and then spatial averaging using (\ref{eq:averaged}), we find (\ref{eq:PandM}),
where the macroscopic electric dipole moment per unit volume, electric
quadrupole moment per unit volume, and magnetic dipole moment per
unit volume are given by 
\begin{align}
\mathscr{P}^{i}(\boldsymbol{x},t)&=\sum_{\boldsymbol{R}}\mathsf{w}(\boldsymbol{x}-\boldsymbol{R})\mu_{\boldsymbol{R}}^{i}(t),\nonumber\\
\mathscr{Q}^{ij}(\boldsymbol{x},t)&=\sum_{\boldsymbol{R}}\mathsf{w}(\boldsymbol{x}-\boldsymbol{R})q_{\boldsymbol{R}}^{ij}(t),\nonumber \\
\mathscr{M}^{i}(\boldsymbol{x},t)&=\sum_{\boldsymbol{R}}\mathsf{w}(\boldsymbol{x}-\boldsymbol{R})\nu_{\boldsymbol{R}}^{i}(t),\label{eq:mac_moments} 
\end{align}
respectively. Since $\varrho_{F}(\boldsymbol{x},t)$ and $\boldsymbol{J}_{F}(\boldsymbol{x},t)$ vanish in the problem at hand, upon implementing (\ref{eq:PandM}) in the macroscopic Maxwell equations, (\ref{eq:macroMaxwell}), $\mathscr{P}^{i}(\boldsymbol{x},t)$, $\mathscr{Q}^{ij}(\boldsymbol{x},t)$, and $\mathscr{M}^{i}(\boldsymbol{x},t)$ serve as the only source terms at this level of analysis. The remaining task is to establish the constitutive relations (\ref{eq:linear_response}).

We can do this by inserting (\ref{eq:linear_response-1}) in (\ref{eq:mac_moments}). The terms that will appear involve
\begin{align}
\sum_{\boldsymbol{R}}\mathsf{w}(\boldsymbol{x}-\boldsymbol{R})L(\boldsymbol{R},\omega),
\label{eq:Lwork}
\end{align}
where here $L(\boldsymbol{R},\omega)$ is one of the macroscopic fields $E^{l}(\boldsymbol{R},\omega)$, $B^{l}(\boldsymbol{R},\omega)$, or $F^{jl}(\boldsymbol{R},\omega)$. To investigate this kind of sum we note that
\begin{align}
&\sum_{\boldsymbol{R}}\mathsf{w}(\boldsymbol{x}-\boldsymbol{R})e^{i\boldsymbol{q}\boldsymbol{\cdot}\boldsymbol{R}}\nonumber\\
&=\int d\boldsymbol{x}'\mathsf{w}(\boldsymbol{x}-\boldsymbol{x}')\left(\sum_{\boldsymbol{R}}\delta(\boldsymbol{x}'-\boldsymbol{R})\right)e^{i\boldsymbol{q}\boldsymbol{\cdot}\boldsymbol{x}'}\nonumber\\
&=\frac{1}{\Omega_{uc}}e^{i\boldsymbol{q}\boldsymbol{\cdot}\boldsymbol{x}}\mathsf{w}(\boldsymbol{q})+\frac{1}{\Omega_{uc}}\sum_{\boldsymbol{G}\neq\boldsymbol{0}}\mathsf{w}(\boldsymbol{q}+\boldsymbol{G})e^{i(\boldsymbol{q}+\boldsymbol{G})\boldsymbol{\cdot}\boldsymbol{x}},\label{eq:sum_work} 
\end{align}
where the $\boldsymbol{G}$ are reciprocal lattice vectors, and we
have used 
\begin{align*}
\sum_{\boldsymbol{R}}\delta(\boldsymbol{x}-\boldsymbol{R})=\frac{1}{\Omega_{uc}}\sum_{\boldsymbol{G}}e^{i\boldsymbol{G}\boldsymbol{\cdot}\boldsymbol{x}}.
\end{align*}
If we choose $\mathsf{w}(\boldsymbol{q})=\mathsf{w}_{I}(\boldsymbol{q})$, then the $\boldsymbol{q}$ that will contribute to $L(\boldsymbol{R},\omega)$ are such that the second term in the final equality of (\ref{eq:sum_work}) rigorously vanishes; from the first term in that expression we see that, since $\mathsf{w}_{I}(\boldsymbol{q})$ acts as a projector, we will have 
\begin{align}
\sum_{\boldsymbol{R}}\mathsf{w}(\boldsymbol{x}-\boldsymbol{R})L(\boldsymbol{R},\omega)=\frac{1}{\Omega_{uc}}L(\boldsymbol{x},\omega),\label{eq:proj_result}
\end{align}
exactly. If we choose $\mathsf{w}(\boldsymbol{q})=\mathsf{w}_{II}(\boldsymbol{q})$, then  there will be corrections to this, since $\mathsf{w}_{II}(\boldsymbol{q})$ does not act as a projector. However, the corrections will be small given that the inequalities (\ref{eq:inequalities}) are assumed to be satisfied, and we can redefine our local field corrections to include them. We then find that (\ref{eq:mac_moments},\ref{eq:linear_response-1},\ref{eq:proj_result}) lead to (\ref{eq:linear_response}), the form of our constitutive relations.

\section{List of response tensors}
\label{Appendix:Tensors}

We here list all the response tensors that were found in this work. The derivation of these response tensors, including the acknowledgment of the assumptions that have been made, and the identification of the quantities they relate, is presented in Section \ref{Section2}--\ref{Section4}. The response tensor $\chi^{il}_{E}(\omega)$ is gauge invariant. For all the other tensors, the portion indicated with a breve accent is gauge invariant.
\begin{widetext}
\begin{align*}
\chi^{il}_E(\omega)= e^2\sum_{mn}f_{nm}\int_{\text{BZ}}\frac{d\boldsymbol{k}}{(2\pi)^3}\frac{\xi^l_{mn}\xi^i_{nm}}{E_{m\boldsymbol{k}}-E_{n\boldsymbol{k}}-\hbar(\omega+i0^+)}.
\end{align*}
\begin{align*}
\gamma^{ijl}(\omega)=\breve{\gamma}^{ijl}(\omega)+\frac{e^2}{4}\sum_{mns}f_{nm}\int_{\text{BZ}}\frac{d\boldsymbol{k}}{(2\pi)^3}\frac{(\xi^l_{mn}\mathcal{W}^j_{ns}+\xi^j_{mn}\mathcal{W}^l_{ns})\xi^i_{sm}+\xi^i_{ns}(\mathcal{W}^j_{sm}\xi^l_{mn}+\mathcal{W}^l_{sm}\xi^j_{mn})}{E_{m\boldsymbol{k}}-E_{n\boldsymbol{k}}-\hbar(\omega+i0^+)},
\end{align*}
where
\begin{align*}
\breve{\gamma}^{ijl}(\omega)\equiv\frac{e^2}{4}\sum_{mn}f_{nm}\int_{\text{BZ}}\frac{d\boldsymbol{k}}{(2\pi)^3}\Big(\mathscr{F}^{jl}_{mn}(\boldsymbol{k},\omega)+\mathscr{F}^{lj}_{mn}(\boldsymbol{k},\omega)\Big)\xi^i_{nm},
\end{align*}
and ${\mathscr{F}}^{jl}_{mn}(\boldsymbol{k},\omega)$ is given by (\ref{scrF}).
\begin{align*}
\chi_{\mathscr{Q}}^{ijl}(\omega)&=\breve{\chi}_{\mathscr{Q}}^{ijl}(\omega)+\frac{e^2}{4}\sum_{mns}f_{nm}\int_{\text{BZ}}\frac{d\boldsymbol{k}}{(2\pi)^3}\Bigg(\frac{\xi^l_{mn}(\xi^i_{ns}\mathcal{W}^j_{sm}+\mathcal{W}^j_{ns}\xi^i_{sm})}{E_{m\boldsymbol{k}}-E_{n\boldsymbol{k}}-\hbar(\omega+i0^+)}+\frac{\xi^l_{mn}(\xi^j_{ns}\mathcal{W}^i_{sm}+\mathcal{W}^i_{ns}\xi^j_{sm})}{E_{m\boldsymbol{k}}-E_{n\boldsymbol{k}}-\hbar(\omega+i0^+)}\Bigg)
\end{align*}
where
\begin{align*}
\breve{\chi}_{\mathscr{Q}}^{ijl}(\omega)\equiv\frac{e^2}{4}\sum_{mns}f_{nm}\int_{\text{BZ}}\frac{d\boldsymbol{k}}{(2\pi)^3}\frac{\xi^l_{mn}(\xi^i_{ns}\xi^j_{sm}+\xi^j_{ns}\xi^i_{sm})}{E_{m\boldsymbol{k}}-E_{n\boldsymbol{k}}-\hbar(\omega+i0^+)}.
\end{align*}
\begin{align*}
\alpha^{il}_{\mathscr{P}}(\omega)=\breve{\alpha}^{il}_{\mathscr{P}}(\omega)+\frac{i\omega e^2}{4c}\epsilon^{lab}\sum_{mns}f_{nm}\int_{\text{BZ}}\frac{d\boldsymbol{k}}{(2\pi)^3}\frac{\xi^i_{ns}\mathcal{W}^a_{sm}\xi^b_{mn}+\xi^b_{mn}\mathcal{W}^a_{ns}\xi^i_{sm}}{E_{m\boldsymbol{k}}-E_{n\boldsymbol{k}}-\hbar(\omega+i0^+)},
\end{align*}
where
\begin{align*}
\breve{\alpha}^{il}_{\mathscr{P}}(\omega)&\equiv\frac{\omega e^2}{4c}\epsilon^{lab}\sum_{mn}f_{nm}\int_{\text{BZ}}\frac{d\boldsymbol{k}}{(2\pi)^3}\frac{\acute{\mathscr{B}}^{ab}_{mn}(\boldsymbol{k},\omega)\xi^i_{nm}}{E_{m\boldsymbol{k}}-E_{n\boldsymbol{k}}-\hbar(\omega+i0^+)},
\end{align*}
and $\acute{\mathscr{B}}^{ab}_{mn}(\boldsymbol{k},\omega)$ is defined in (\ref{breveB}).
\begin{align*}
\alpha^{li}_{\mathscr{M}}(\omega)=\breve{\alpha}^{li}_{\mathscr{M}}(\omega)-\frac{i\omega e^2}{4c}\epsilon^{iab}\sum_{mns}f_{nm}\int_{\text{BZ}}\frac{d\boldsymbol{k}}{(2\pi)^3}\frac{\xi^l_{mn}\mathcal{W}^a_{ns}\xi^b_{sm}+\xi^b_{ns}\mathcal{W}^a_{sm}\xi^l_{mn}}{E_{m\boldsymbol{k}}-E_{n\boldsymbol{k}}-\hbar(\omega+i0^+)},
\end{align*}
where
\begin{align*}
\breve{\alpha}^{li}_{\mathscr{M}}(\omega)&\equiv\frac{\omega e^2}{4c}\epsilon^{iab}\sum_{mn}f_{nm}\int_{\text{BZ}}\frac{d\boldsymbol{k}}{(2\pi)^3}\frac{1}{E_{m\boldsymbol{k}}-E_{n\boldsymbol{k}}-\hbar(\omega+i0^+)}\\
&\qquad\times\Bigg\{2\frac{\partial_b(E_{m\boldsymbol{k}}+E_{n\boldsymbol{k}})}{E_{m\boldsymbol{k}}-E_{n\boldsymbol{k}}}\xi^a_{nm}\xi^l_{mn}+i\sum_{s}\frac{E_{s\boldsymbol{k}}-E_{m\boldsymbol{k}}}{E_{m\boldsymbol{k}}-E_{n\boldsymbol{k}}}\xi^a_{ns}\xi^b_{sm}\xi^l_{mn}+i\sum_{s}\frac{E_{n\boldsymbol{k}}-E_{s\boldsymbol{k}}}{E_{m\boldsymbol{k}}-E_{n\boldsymbol{k}}}\xi^b_{ns}\xi^a_{sm}\xi^l_{mn}\Bigg\}.
\end{align*}

We have previously \cite{Mahon2020} found the OMP tensor to be of the form
\begin{align}
\alpha^{il}=\alpha^{il}_{\text{G}}+\delta^{il}\alpha_{\text{CS}},
\label{OMP}
\end{align}
where
\begin{align}
{\alpha}^{il}_{\text{G}}&=\frac{e^2}{\hbar c}\epsilon^{lab}\int_{\text{BZ}}\frac{d\boldsymbol{k}}{(2\pi)^3}\Bigg\{-\sum_{cv}\frac{\partial_b( E_{c\boldsymbol{k}}+E_{v\boldsymbol{k}})}{E_{v\boldsymbol{k}}-E_{c\boldsymbol{k}}}\text{Re}\big[\left(\partial_av|c\right)\left(c|\partial_iv\right)\big]-\sum_{cvv'}\frac{E_{v\boldsymbol{k}}-E_{v'\boldsymbol{k}}}{E_{v\boldsymbol{k}}-E_{c\boldsymbol{k}}}\text{Re}\big[\left(\partial_bv|v'\right)\left(\partial_av'|c\right)\left(c|\partial_iv\right)\big]\nonumber\\
&\quad\qquad\qquad\qquad\qquad\qquad+\sum_{cc'v}\frac{E_{c\boldsymbol{k}}-E_{c'\boldsymbol{k}}}{E_{v\boldsymbol{k}}-E_{c\boldsymbol{k}}}\text{Re}\big[\left(\partial_bv|c'\right)\left(c'|\partial_ac\right)\left(c|\partial_iv\right)\big]\Bigg\},
\label{alphaG}
\end{align}
and
\begin{align}
\alpha_{\text{CS}}=-\frac{e^2}{2\hbar c}\epsilon^{abd}\int_{\text{BZ}}\frac{d\boldsymbol{k}}{(2\pi)^3}\Bigg[\left(\sum_{vv'}\xi^a_{vv'}\partial_b\xi^d_{v'v}-\frac{2i}{3}\sum_{vv'v_1}\xi^a_{vv'}\xi^b_{v'v_1}\xi^d_{v_1v}\right)+\sum_{vv'}(\partial_b\mathcal{W}^a_{vv'})\mathcal{W}^d_{v'v}-\frac{2i}{3}\sum_{vv'v_1}\mathcal{W}^a_{vv'}\mathcal{W}^b_{v'v_1}\mathcal{W}^d_{v_1v}\Bigg],
\label{alphaCS}
\end{align}
where $\alpha^{il}_{\text{G}}$ is gauge invariant and $\alpha_{\text{CS}}$ is not.

\section{Gauge invariance of induced first-order macroscopic current density}
\label{Appendix:InducedCurrent}
We begin by separating (\ref{inducedCurrent}) into a sum of gauge-invariant and gauge-dependent terms. We then collect the gauge-invariant terms, i.e., $\breve{\chi}$ contributions, into $[\ldots]$. We find
\begin{align*}
J^{i(1)}(\boldsymbol{x},\omega)&=-i\omega\chi_E^{il}(\omega)E^l(\boldsymbol{x},\omega)-i\omega\gamma^{ijl}(\omega)F^{jl}(\boldsymbol{x},\omega)-i\omega\big(\alpha^{il}_{\text{G}}+\alpha_P^{il}(\omega)\big)B^l(\boldsymbol{x},\omega) \\
&\quad\qquad\qquad\qquad+i\omega\chi_{\mathscr{Q}}^{ijl}(\omega)\frac{\partial E^l(\boldsymbol{x},\omega)}{\partial x^j}+c\epsilon^{iab}\big(\alpha^{lb}_{\text{G}}+\alpha_M^{lb}(\omega)\big)\frac{\partial E^l(\boldsymbol{x},\omega)}{\partial x^a}\\
&=\big[\ldots\big]-i\frac{e^2}{2}\sum_{mns}f_{nm}\int_{\text{BZ}}\frac{d\boldsymbol{k}}{(2\pi)^3}\Bigg\{\omega\frac{\xi^l_{mn}\mathcal{W}^j_{ns}\xi^i_{sm}+\xi^i_{ns}\mathcal{W}^j_{sm}\xi^l_{mn}}{E_{m\boldsymbol{k}}-E_{n\boldsymbol{k}}-\hbar(\omega+i0^+)}F^{jl}(\boldsymbol{x},\omega) \\
&\qquad\qquad\qquad\qquad+i\omega\frac{\epsilon^{lab}}{2\hbar c}\frac{\hbar\omega}{E_{m\boldsymbol{k}}-E_{n\boldsymbol{k}}-\hbar(\omega+i0^+)}\left(\xi^i_{ns}\mathcal{W}^a_{sm}\xi^b_{mn}+\xi^b_{mn}\mathcal{W}^a_{ns}\xi^i_{sm}\right)B^l(\boldsymbol{x},\omega) \\
&\qquad\qquad\qquad\qquad-\frac{\omega}{2}\left(\frac{\xi^i_{ns}\mathcal{W}^j_{sm}\xi^l_{mn}+\xi^l_{mn}\mathcal{W}^j_{ns}\xi^i_{sm}}{E_{m\boldsymbol{k}}-E_{n\boldsymbol{k}}-\hbar(\omega+i0^+)}+\frac{\xi^j_{ns}\mathcal{W}^i_{sm}\xi^l_{mn}+\xi^l_{mn}\mathcal{W}^i_{ns}\xi^j_{sm}}{E_{m\boldsymbol{k}}-E_{n\boldsymbol{k}}-\hbar(\omega+i0^+)}\right)\frac{\partial E^l(\boldsymbol{x},\omega)}{\partial x^j} \\
&\qquad\qquad\qquad\qquad+c\epsilon^{iab}\frac{\epsilon^{bcd}}{2\hbar c}\frac{\hbar\omega}{E_{m\boldsymbol{k}}-E_{n\boldsymbol{k}}-\hbar(\omega+i0^+)}\left(\xi^l_{mn}\mathcal{W}^c_{ns}\xi^d_{sm}+\xi^d_{ns}\mathcal{W}^c_{sm}\xi^l_{mn}\right)\frac{\partial E^l(\boldsymbol{x},\omega)}{\partial x^a}\Bigg\} \\
&=\big[\ldots\big]-i\frac{e^2}{2}\sum_{mns}f_{nm}\int_{\text{BZ}}\frac{d\boldsymbol{k}}{(2\pi)^3}\Bigg\{\frac{\omega}{2}\frac{\xi^l_{mn}\mathcal{W}^j_{ns}\xi^i_{sm}+\xi^i_{ns}\mathcal{W}^j_{sm}\xi^l_{mn}}{E_{m\boldsymbol{k}}-E_{n\boldsymbol{k}}-\hbar(\omega+i0^+)}\left[\frac{\partial E^l(\boldsymbol{x},\omega)}{\partial x^j}+\frac{\partial E^j(\boldsymbol{x},\omega)}{\partial x^l}\right] \\
&\qquad\qquad\qquad\qquad+\epsilon^{lab}\frac{\omega}{2}\frac{\xi^i_{ns}\mathcal{W}^a_{sm}\xi^b_{mn}+\xi^b_{mn}\mathcal{W}^a_{ns}\xi^i_{sm}}{E_{m\boldsymbol{k}}-E_{n\boldsymbol{k}}-\hbar(\omega+i0^+)}\epsilon^{lcd}\frac{\partial E^d(\boldsymbol{x},\omega)}{\partial x^c} \\
&\qquad\qquad\qquad\qquad-\frac{\omega}{2}\left(\frac{\xi^i_{ns}\mathcal{W}^j_{sm}\xi^l_{mn}+\xi^l_{mn}\mathcal{W}^j_{ns}\xi^i_{sm}}{E_{m\boldsymbol{k}}-E_{n\boldsymbol{k}}-\hbar(\omega+i0^+)}+\frac{\xi^j_{ns}\mathcal{W}^i_{sm}\xi^l_{mn}+\xi^l_{mn}\mathcal{W}^i_{ns}\xi^j_{sm}}{E_{m\boldsymbol{k}}-E_{n\boldsymbol{k}}-\hbar(\omega+i0^+)}\right)\frac{\partial E^l(\boldsymbol{x},\omega)}{\partial x^j} \\
&\qquad\qquad\qquad\qquad+\epsilon^{iab}\epsilon^{bcd}\frac{\omega}{2}\frac{\xi^l_{mn}\mathcal{W}^c_{ns}\xi^d_{sm}+\xi^d_{ns}\mathcal{W}^c_{sm}\xi^l_{mn}}{E_{m\boldsymbol{k}}-E_{n\boldsymbol{k}}-\hbar(\omega+i0^+)}\frac{\partial E^l(\boldsymbol{x},\omega)}{\partial x^a}\Bigg\} 
\end{align*}
\begin{align*}
&=\big[\ldots\big]-i\frac{e^2}{2}\sum_{mns}f_{nm}\int_{\text{BZ}}\frac{d\boldsymbol{k}}{(2\pi)^3}\Bigg\{\frac{\omega}{2}\frac{\xi^l_{mn}\mathcal{W}^j_{ns}\xi^i_{sm}+\xi^i_{ns}\mathcal{W}^j_{sm}\xi^l_{mn}}{E_{m\boldsymbol{k}}-E_{n\boldsymbol{k}}-\hbar(\omega+i0^+)}\frac{\partial E^j(\boldsymbol{x},\omega)}{\partial x^l} \\
&\qquad\qquad\qquad\qquad+\frac{\omega}{2}\frac{\xi^i_{ns}\mathcal{W}^a_{sm}\xi^b_{mn}+\xi^b_{mn}\mathcal{W}^a_{ns}\xi^i_{sm}}{E_{m\boldsymbol{k}}-E_{n\boldsymbol{k}}-\hbar(\omega+i0^+)}\left[\frac{\partial E^b(\boldsymbol{x},\omega)}{\partial x^a}-\frac{\partial E^a(\boldsymbol{x},\omega)}{\partial x^b}\right] \\
&\qquad\qquad\qquad\qquad-\frac{\omega}{2}\frac{\xi^j_{ns}\mathcal{W}^i_{sm}\xi^l_{mn}+\xi^l_{mn}\mathcal{W}^i_{ns}\xi^j_{sm}}{E_{m\boldsymbol{k}}-E_{n\boldsymbol{k}}-\hbar(\omega+i0^+)}\frac{\partial E^l(\boldsymbol{x},\omega)}{\partial x^j} \\
&\qquad\qquad\qquad\qquad+\frac{\omega}{2}\left(\frac{\xi^l_{mn}\mathcal{W}^i_{ns}\xi^a_{sm}+\xi^a_{ns}\mathcal{W}^i_{sm}\xi^l_{mn}}{E_{m\boldsymbol{k}}-E_{n\boldsymbol{k}}-\hbar(\omega+i0^+)}-\frac{\xi^l_{mn}\mathcal{W}^a_{ns}\xi^i_{sm}+\xi^i_{ns}\mathcal{W}^a_{sm}\xi^l_{mn}}{E_{m\boldsymbol{k}}-E_{n\boldsymbol{k}}-\hbar(\omega+i0^+)}\right)\frac{\partial E^l(\boldsymbol{x},\omega)}{\partial x^a}\Bigg\} \\
&=\big[\ldots\big]-i\frac{e^2}{2}\sum_{mns}f_{nm}\int_{\text{BZ}}\frac{d\boldsymbol{k}}{(2\pi)^3}\Bigg\{\frac{\omega}{2}\frac{\xi^l_{mn}\mathcal{W}^j_{ns}\xi^i_{sm}+\xi^i_{ns}\mathcal{W}^j_{sm}\xi^l_{mn}}{E_{m\boldsymbol{k}}-E_{n\boldsymbol{k}}-\hbar(\omega+i0^+)}\frac{\partial E^j(\boldsymbol{x},\omega)}{\partial x^l} \\
&\qquad\qquad\qquad\qquad+\frac{\omega}{2}\frac{\xi^i_{ns}\mathcal{W}^a_{sm}\xi^b_{mn}+\xi^b_{mn}\mathcal{W}^a_{ns}\xi^i_{sm}}{E_{m\boldsymbol{k}}-E_{n\boldsymbol{k}}-\hbar(\omega+i0^+)}\left[\frac{\partial E^b(\boldsymbol{x},\omega)}{\partial x^a}-\frac{\partial E^a(\boldsymbol{x},\omega)}{\partial x^b}\right] \\
&\qquad\qquad\qquad\qquad+\frac{\omega}{2}\left(-\frac{\xi^l_{mn}\mathcal{W}^a_{ns}\xi^i_{sm}+\xi^i_{ns}\mathcal{W}^a_{sm}\xi^l_{mn}}{E_{m\boldsymbol{k}}-E_{n\boldsymbol{k}}-\hbar(\omega+i0^+)}\right)\frac{\partial E^l(\boldsymbol{x},\omega)}{\partial x^a}\Bigg\} \\
&=\big[\ldots\big],
\end{align*}
where in the above we have used the identity $\epsilon^{lab}\epsilon^{lcd}=\delta^{ac}\delta^{bd}-\delta^{ad}\delta^{bc}$. As the $[\ldots]$ term contains only the gauge-invariant contributions of the response tensors in (\ref{inducedCurrent}), we arrive at (\ref{inducedCurrentDensity}).

\section{Gauge invariance of induced first-order macroscopic charge density}
\label{Appendix:InducedCharge}
We begin by separating (\ref{inducedCharge}) into a sum of gauge-invariant and gauge-dependent terms. We then collect the gauge-invariant terms, i.e., $\breve{\chi}$ contributions, into $[\ldots]$. We find
\begin{align*}
\varrho^{(1)}(\boldsymbol{x},\omega)&=-\Bigg(\chi_E^{al}(\omega)\frac{\partial E^l(\boldsymbol{x},\omega)}{\partial x^a}+\gamma^{ajl}(\omega)\frac{\partial F^{jl}(\boldsymbol{x},\omega)}{\partial x^a}+\big(\alpha_{\text{G}}^{al}+\alpha_P^{al}(\omega)\big)\frac{\partial B^l(\boldsymbol{x},\omega)}{\partial x^a}-\chi_{\mathscr{Q}}^{ajl}(\omega)\frac{\partial^2 E^l(\boldsymbol{x},\omega)}{\partial x^a\partial x^j}\Bigg) \\
&=-\big[\ldots\big]-\frac{e^2}{2}\sum_{mns}f_{nm}\int_{\text{BZ}}\frac{d\boldsymbol{k}}{(2\pi)^3}\Bigg\{\frac{\xi^l_{mn}\mathcal{W}^j_{ns}\xi^a_{sm}+\xi^a_{ns}\mathcal{W}^j_{sm}\xi^l_{mn}}{E_{m\boldsymbol{k}}-E_{n\boldsymbol{k}}-\hbar(\omega+i0^+)}\frac{\partial F^{jl}(\boldsymbol{x},\omega)}{\partial x^a} \\
&\quad\qquad\qquad\qquad+\frac{\epsilon^{ldb}}{2 c}\frac{i\omega}{E_{m\boldsymbol{k}}-E_{n\boldsymbol{k}}-\hbar(\omega+i0^+)}\xi^a_{ns}\mathcal{W}^d_{sm}\xi^b_{mn}+\xi^b_{mn}\mathcal{W}^d_{ns}\xi^a_{sm}\frac{\partial B^l(\boldsymbol{x},\omega)}{\partial x^a} \\
&\quad\qquad\qquad\qquad-\frac{\xi^a_{ns}\mathcal{W}^j_{sm}\xi^l_{mn}+\xi^l_{mn}\mathcal{W}^j_{ns}\xi^a_{sm}}{E_{m\boldsymbol{k}}-E_{n\boldsymbol{k}}-\hbar(\omega+i0^+)}\frac{\partial^2 E^l(\boldsymbol{x},\omega)}{\partial x^a\partial x^j}\Bigg\} \\
&=-\big[\ldots\big]-\frac{e^2}{4}\sum_{mns}f_{nm}\int_{\text{BZ}}\frac{d\boldsymbol{k}}{(2\pi)^3}\Bigg\{\frac{\xi^l_{mn}\mathcal{W}^j_{ns}\xi^a_{sm}+\xi^a_{ns}\mathcal{W}^j_{sm}\xi^l_{mn}}{E_{m\boldsymbol{k}}-E_{n\boldsymbol{k}}-\hbar(\omega+i0^+)}\left[-\frac{\partial^2 E^l(\boldsymbol{x},\omega)}{\partial x^a\partial x^j}+\frac{\partial^2 E^j(\boldsymbol{x},\omega)}{\partial x^a\partial x^l}\right] \\
&\quad\qquad\qquad\qquad+\epsilon^{ldb}\epsilon^{lce}\frac{\xi^a_{ns}\mathcal{W}^d_{sm}\xi^b_{mn}+\xi^b_{mn}\mathcal{W}^d_{ns}\xi^a_{sm}}{E_{m\boldsymbol{k}}-E_{n\boldsymbol{k}}-\hbar(\omega+i0^+)}\frac{\partial^2E^e(\boldsymbol{x},\omega)}{\partial x^c\partial x^a}\Bigg\} \\
&=-\big[\ldots\big]-\frac{e^2}{4}\sum_{mns}f_{nm}\int_{\text{BZ}}\frac{d\boldsymbol{k}}{(2\pi)^3}\Bigg\{\frac{\xi^b_{mn}\mathcal{W}^d_{ns}\xi^a_{sm}+\xi^a_{ns}\mathcal{W}^d_{sm}\xi^b_{mn}}{E_{m\boldsymbol{k}}-E_{n\boldsymbol{k}}-\hbar(\omega+i0^+)}\left[-\frac{\partial^2 E^b(\boldsymbol{x},\omega)}{\partial x^a\partial x^d}+\frac{\partial^2 E^d(\boldsymbol{x},\omega)}{\partial x^a\partial x^b}\right] \\
&\quad\qquad\qquad\qquad+\frac{\xi^a_{ns}\mathcal{W}^d_{sm}\xi^b_{mn}+\xi^b_{mn}\mathcal{W}^d_{ns}\xi^a_{sm}}{E_{m\boldsymbol{k}}-E_{n\boldsymbol{k}}-\hbar(\omega+i0^+)}\left[\frac{\partial^2E^b(\boldsymbol{x},\omega)}{\partial x^d\partial x^a}-\frac{\partial^2E^d(\boldsymbol{x},\omega)}{\partial x^b\partial x^a}\right] \Bigg\} \\
&=-\big[\ldots\big].
\end{align*}
As the $[\ldots]$ term contains only the gauge-invariant contributions of the response tensors in (\ref{inducedCharge}), we arrive at (\ref{inducedChargeDensity}).

\section{Linear-in-$\boldsymbol{q}$ contribution to optical conductivity}
\label{Appendix:Conductivity}
In past work, Malashevich and Souza \cite{Souza2010} introduce $B^{ab(\text{orb})}_{nm}(\boldsymbol{k})$, which we call $\mathcal{B}^{ab}_{nm}(\boldsymbol{k})$. We arrive at (\ref{calB}) from Eq.~(36) presented there in the following way. We begin by implementing the definition of the non-Abelian Berry connection
\begin{align*}
i\ket{\partial_an\boldsymbol{k}}=\sum_{n'}\xi^{a}_{n'n}(\boldsymbol{k})\ket{n'\boldsymbol{k}},
\end{align*}
and use the identity 
\begin{align*}
\partial_aH_{\boldsymbol{k}}\ket{n\boldsymbol{k}}=\partial_aE_{n\boldsymbol{k}}\ket{n\boldsymbol{k}}+(E_{n\boldsymbol{k}}-H_{\boldsymbol{k}})\ket{\partial_an\boldsymbol{k}}.
\end{align*}
Implementing this, and $H_{\boldsymbol{k}}\ket{n\boldsymbol{k}}=E_{n\boldsymbol{k}}\ket{n\boldsymbol{k}}$, we arrive at (\ref{calB}).
\end{widetext}

\bibliographystyle{apsrev4-1}
\bibliography{acResponse_Updated}

\end{document}